%% file: cost1014.tex
\documentclass[preprint]{aastex}
\usepackage{natbib}
\usepackage{multirow}
\usepackage[caption=false]{subfig}
\include{defs}
\usepackage{graphicx}
\usepackage{amsmath}
\usepackage[normalem]{ulem}
\usepackage{ mathrsfs }

\usepackage{threeparttable}

\begin{document}
\title{Denser Sampling of the Rosette Nebula with Faraday Rotation Measurements: Improved Estimates of Magnetic Fields in \HII Regions}
\author{Allison H. Costa\altaffilmark{1}, Steven R. Spangler\altaffilmark{1}, Joseph R. Sink\altaffilmark{1}, Shea Brown\altaffilmark{1}, and Sui Ann Mao\altaffilmark{2}}
\altaffiltext{1}{Department of Physics and Astronomy, University of Iowa, Iowa City, Iowa 52242}
\altaffiltext{2}{Max Planck Institute for Radio Astronomy, Bonn, Germany}

\begin{abstract}
We report Faraday rotation measurements of 11 extragalactic radio sources with lines of sight through the Rosette Nebula, a prominent \HII region associated with the star cluster NGC 2244. 
The goal of these measurements is to better determine the strength and structure of the magnetic field in the nebula.  We calculate the rotation measure (RM) through two methods, a least-squares fit to \chilam~and Rotation Measure Synthesis. In conjunction with our results from \citet{Savage:2013}, we find an excess RM due to the shell of the nebula of +40 to +1200 \radm~above a background RM of +147~\radm. We discuss two forms of a simple shell model intended to reproduce the magnitude of the observed RM as a function of distance from the center of the Rosette Nebula. The models represent different physical situations for the magnetic field within the shell of the nebula. The first assumes that there is an increase in the magnetic field strength and plasma density at the outer radius of the \HII region, such as would be produced by a strong magnetohydrodynamic shock wave. The second model assumes that any increase in the RM is due solely to an increase in the density, and the Galactic magnetic field is unaffected in the shell. We employ a Bayesian analysis to compare the two forms of the model. The results of this analysis were inconclusive, although the model without amplification of the interstellar magnetic field is weakly favored.

\end{abstract}


\section{Introduction}\label{intro}
Throughout the main sequence lifetimes of O and B stars, these massive stars modify the surrounding matter by photoionizing it, creating an \HII region, and through their stellar winds. The stellar wind expands out into the interstellar medium (ISM), sweeping up material, and inflating a bubble of hot, ionized gas. The \citet{Weaver:1977} solution for a stellar bubble inflated by the stellar wind of a single star consists of four regions. These regions are (see Figure 1 of \citealt{Weaver:1977}) (a) the inner region closest to the star with the hypersonic stellar wind, (b) a bubble of hot, low density gas, (c) an annular shell of swept up shocked ISM gas, which may constitute part or all of the observed \HII region, and (d) the ambient ISM exterior to the bubble. In Figure 1 of \citet{Weaver:1977}, region (c) constitutes an \HII region. A further diagram of an \HII region is shown in Figure 3 of \citet{Weaver:1977}. \HII regions are plasmas, and magnetic fields affect the dynamics of \HII regions and stellar bubbles through magnetic pressure and magnetic tension. In magnetohydrodynamic (MHD) simulations, magnetic fields can elongate the shells of stellar bubbles preferentially in the direction of the field lines and may thicken the shell perpendicular to the magnetic field, altering the shape of the stellar bubble \citep{Ferriere:1991,Stil:2009}. Possible observations of the elongation of young bubbles with respect to magnetic fields are discussed in \citet{Pavel:2012}. However, magnetic field properties are difficult to measure in \HII regions. 

The goal of this research is to understand how the general interstellar magnetic field, B$_{\textrm{{\scriptsize ISM}}}$, is modified in the interior of the \HII region. The technique we employ to investigate the role of magnetic fields in stellar bubbles is Faraday rotation. Faraday rotation is the rotation of the plane of polarization of a radio wave as it passes through a plasma that contains a magnetic field and is described by
\begin{equation}
\chi=\chi_{0}+\left[\left(\frac{e^{3}}{2\pi m_{e}^{2}c^{4}}\right)\int_{\text{source}}^{\text{observer}} {n_e\mathbf{B}\cdot \mathbf{ds}}\right]\lambda^{2},
\label{RM1}
\end{equation}
where $\chi$ is the polarization position angle of a radio wave after propagating through a plasma, $\chi_{0}$ is the intrinsic polarization position angle of the radio wave before propagating through a plasma, $e$, $m_{e}$, and $c$ are the usual fundamental physical constants, $n_{e}$ is the electron density of the plasma, \textbf{B} is the vector magnetic field, \textbf{ds} is a vector increment of the path length from the source to the observer, and $\lambda$ is the wavelength. The quantity in the square brackets in Equation (\ref{RM1}) is termed the rotation measure (RM) and can be written in convenient units as
\begin{equation}
\textnormal{RM}=0.81\int
n_{e} (\text{cm$^{-3}$}) \mathbf{B}(\mu\text{G})\cdot \mathbf{ds} \text{ (pc)  rad m$^{-2}$}
\label{RMSI}
\end{equation}
\citep{Minter:1996}, where the term in the parentheses in Equation (\ref{RM1}) equals 0.81 in these units. 
To obtain information on the magnetic field, the electron density needs to be independently determined since the integrand in Equation (\ref{RMSI}) is the product of n$_{e}$ and \textbf{B}. Such independent measurements are provided by a number of techniques such as thermal radio emission (used in this paper), intensity of radio recombination lines, or pulsar dispersion.

The Rosette Nebula is a good candidate  for this line of research, as it is an \HII region with an obvious shell structure and central cavity. \citet{Menon:1962} determined that the Rosette Nebula is ionization bounded from radio continuum observations, which was later confirmed by \citet{Celnik:1983,Celnik:1985,Celnik:1986}. \citet{Celnik:1985} determined the inner and outer radii of the shell and the electron density from radio continuum observations at 1410 and 4750 MHz with the 100 m telescope of the Max Planck Institut f\"ur Radioastronomie at Effelsberg. 

The star cluster responsible for the \HII region is the OB stellar association NGC 2244. The nominal center that we adopt for the center of the \HII region is the center of NGC 2244, R.A.(J2000) = 06$^{\textnormal{h}}$ 31$^{\textnormal{m}}$ 55$^{\textnormal{s}}$, Dec.(J2000) = 04$^{\circ}$ 56$'$ 34$''$ ($l$ =206.5, $b$ = --2.1) \citep{Berghofer:2002}. The nebula is located 1600 parsecs away \citep{Roman:2008}.  The age of NGC 2244 is less than 4 Myr old \citep{Perez:1989}, and there are 7 O type stars within the association \citep{Park:2002,Roman:2008,Wang:2008}. The stellar winds of these stars are believed to have inflated a bubble of hot ionized gas around the star cluster, which provides the environment for our Faraday rotation study. The mass loss rates of these O stars have been estimated to be of order $\dot{\textrm{M}}$ $\sim$ \(10^{-6} \) \Msun~yr$^{-1}$ \citep{Howarth:1989}, terminal wind velocities of order v$_{\textrm{{\scriptsize term}}}$ $\sim$ 3000 \kms~ \citep{Chlebowski:1991}, and  wind luminosities L$_w$ $\sim$ 10$^{36}$ ergs s$^{-1}$, where \(\textrm{L}_{w}=\frac{1}{2}\dot{\textrm{M}}\textrm{v}_{\textrm{{\scriptsize term}}}^2 \).  However, recent studies by \citet{Bouret:2005} and \citet{Mokiem:2007} have shown that the mass loss rates of stars may be over estimated by a factor of 3--5 due to clumping in the winds. These lower mass loss rates and wind luminosities would modify the expected evolution of stellar bubbles in general, including the Rosette.

\subsection{Previous Results on Faraday Rotation through \HII Regions\label{sec:prev}}
In \citet{Savage:2013}, we investigated the role of magnetic fields in \HII regions with polarimetric observations of extra galactic radio sources whose lines of sight pass through or close to the Rosette Nebula. We made observations of 23 background radio sources using the Karl G. Jansky Very Large Array. The background radio sources were selected from the National Radio Astronomy Observatory VLA Sky Survey (NVSS) \citep{Condon:1998}. Twelve sources had lines of sight within 1\ddeg~of the nominal center of the nebula, and the remaining 11 sources had lines of sight that passed through an annulus of 1\ddeg--2\ddeg~of the center.

\citet{Savage:2013} measured a background RM due to the Galactic plane in this region of the sky of +147 \radm~and an excess RM of +50 to +750 \radm~due to the shell of the Rosette Nebula \HII region (Table 3 in \citet{Savage:2013}). We employed a physically motivated shell model developed by \citet{Whiting:2009} to reproduce the magnitude and sign of the RM observed in the shell as a function of distance from the center of the nebula (Section 4.1 of \citet{Savage:2013}). The simple shell model assumes spherical symmetry and that the electron density, $n_{e}$, in the shell is an independently determined quantity. The parameters for the inner and outer radii of the shell and the electron density were adopted from Model I of \citet{Celnik:1985}, where it is assumed that the electron density is uniform within an annulus with outer radius of R$_{0}$ and inner radius of R$_{1}$. For a strong shock, the shell model predicts that the largest value of the RM should be near the outer radius of the shell. This ``rotation measure limb brightening'' is due to the MHD shock that is conjectured to define the outer radius, R$_0$, of the bubble. This shock strengthens the interstellar magnetic field and ``refracts'' it into the shock plane (see discussion in Section \ref{sec:emmodels} below).

An assumption of the \citet{Whiting:2009} model, employed in \citet{Savage:2013}, is that the entire Rosette Nebula \HII region  consists of a Weaver-style bubble, with the annular shell thickened by a process such as that suggested by \citet{Breitschwerdt:1988}. 
This model may not be strictly correct.  However, we will employ it in this paper as a simple representation of a class of models in which the magnetic field is amplified at the interface with the HII region.

\citet{Harvey:2011} conducted a Faraday rotation study of five Galactic \HII regions in which they estimated the electron density and the line of sight component of the magnetic field and did not find an amplification of the general interstellar magnetic field due to the shell. In the report of \citet{Harvey:2011},  the increase in the RM is attributed entirely to an increase in the density of the shell. \citet{Savage:2013} employed two simple models for an interstellar shell, one due to a wind-driven stellar bubble as well as one in which there is no change in the interstellar magnetic field, to investigate which physical situation better describes the magnetic field within the shell of the Rosette Nebula. \citet{Savage:2013} argued that the case with an amplified magnetic field in the shell better described the observed magnitude and spatial distribution of the RM from the center of the Rosette Nebula, and suggested that ``RM limb brightening'' was present in the data. In this case, the $|$RM$|$ values are largest near the edge of the shell but then decrease abruptly outside the outer shock front as well as downstream of the contact discontinuity. However, this statement was more a suggestion than a conclusion due to the relatively small number of lines of sight probed within the shell. \citet{Savage:2013} concluded that either of these models was viable as a simplified description of the shell structure.

Recently, \citet{Purcell:2015} performed a Faraday rotation and radio polarization study of the Gum nebula. They suggest that the Gum nebula is not a supernova remnant but is an \HII region surrounding a wind-blown bubble. The ionized shell model they employ has similarities to the one employed by \citet{Savage:2013}. The model of \citet{Purcell:2015} reduces to \citet{Savage:2013} when the nebula subtends a small angle. \citet{Purcell:2015} perform a multi-parameter analysis to simultaneously determine the electron density, shell thickness, filling factor, and the magnetic field instead of assuming fixed parameters, as is done in this paper and \citet{Savage:2013}. 

In this paper, we report and discuss additional data taken to clarify the structure of the Rosette Nebula bubble. We performed polarimetric observations of 11 additional sources observed through the shell of the Rosette Nebula. These sources produce values of RM along 15 lines of sight that are added to those reported in \citet{Savage:2013}. The combined and enlarged data set permits a more detailed study of the magnetic field within the Rosette Nebula bubble. The outline of this paper is as follows. Section \ref{observations} outlines the instrumental configuration and the observations performed. Section \ref{dataredux} describes the data reduction process, and the methods for determining the RM values.  In Section \ref{sec:res}, we report the results of the RM analysis and Section \ref{sec:comp} compares the two methods of determining the RM. Section \ref{section:shells} describes the shell models, and in Section \ref{sec:bay}, we discuss the Bayesian statistical analysis employed to compare the two shell models. In Section \ref{sec:joe}, we modify our shell models to accommodate the inhomogeneity of the nebula. Finally, Section \ref{sec:con} contains our conclusions and summary.

\section{Observations}\label{observations}
\begin{figure}[htb!]
\centering
\includegraphics[scale=0.4]{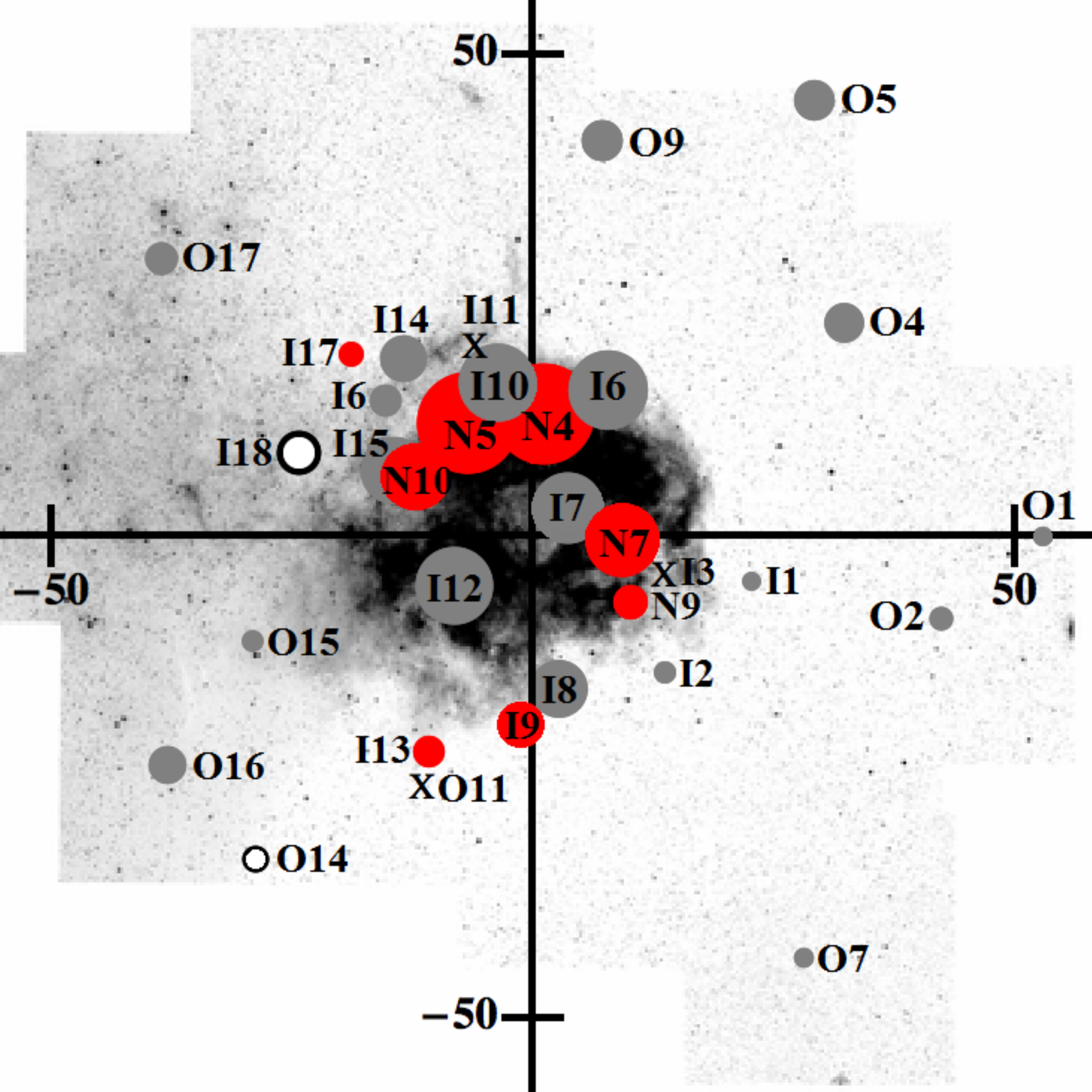}
\caption[Mosaic of Rosette Nebula with new lines of sight]{Palomar Sky Survey Mosaic of the Rosette Nebula with lines of sight probed by the present study and \citet{Savage:2013}. Measurements are represented by circles scaled to $\sqrt{|RM|}$. The red circles represent the new lines of sight observed in this work, and the gray circles are from our previous work (see \citet{Savage:2013}, Figure 1). The `X's are depolarized sources, and the open circles are sources with negative RM values. All other RM values obtained are positive.}
\label{fig:rosettenewlos}
\end{figure}

We observed our target sources with the Jansky Very Large Array (VLA)\footnote{\footnotesize{The Very Large Array is an instrument of the National Radio Astronomy
Observatory. The NRAO is a facility of the National Science Foundation,
operated under cooperative agreement with Associated Universities, Inc.}} radio telescope in C configuration on the 5th of February 2012.
Table \ref{tab:logofobs} lists the details of the observation. We selected 11 extragalactic radio sources with lines of sight within 1\ddeg~of the nominal center of the Rosette Nebula to probe radii within the shell of the nebula. The sources are listed in Table \ref{tab:sources} with their right ascension ($\alpha$), declination ($\delta$), Galactic longitude ($l$), Galactic latitude ($b$), the impact parameter of the line of sight from the center of the nebula ($\xi$), and the peak values of Stokes I intensity at 4.3 GHz. Figure \ref{fig:rosettenewlos} shows the positions of the sources from this work as well as those of \citet{Savage:2013}, plotted over a mosaic of the Rosette Nebula compiled from the Second Palomar Observatory Sky Survey\footnote{The Second Palomar Observatory Sky Survey (POSS-II) was made by the California Institute of Technology with funds from the National Science Foundation, the National Geographic Society, the Sloan Foundation, the Samuel Oschin Foundation, and the Eastman Kodak Corporation. The STScI Digitized Sky Survey can be found at http$://$stdatu.stsci.edu$/$cgi$-$bin$/$dss$_{-}$form.}. 

In addition to the 11 program sources in Table \ref{tab:sources}, we observed J0632$+$1022, J0643$+$0857, and 3C138 as calibrators. The flux density and polarization position angle calibrator for these observations is 3C138. The primary calibrator, J0632$+$1022, was used to determine the complex gain of the antennas as a function of time, as well as the instrumental polarization parameters (D factors). Similarly, we observed J0643$+$0857 as a secondary gain calibrator to independently determine the instrumental polarization and confirm the solutions for the D factors. All sources were observed in a single eight-hour observing run. Five-minute observations of the targets were interleaved with observations of J0632$+$1022 and J0643$+$0857, which guarantees that the calibrators have sufficient parallactic angle coverage to solve for the D terms.

\begin{table}[p]
\centering
\begin{threeparttable}
\caption{Log of Observations \label{tab:logofobs}}

\begin{tabular}{p{0.6\linewidth}
				p{0.25\linewidth}}
\hline
Date of Observations & February 5, 2012\\

Duration of Observing Session (h) & 7.6\\

Frequencies of Observation\tnote{a}  (GHz) & 4.720; 7.250\\

Number of Frequency Channels per IF & 512\\

Channel Width (MHz) & 2 \\

VLA array & C\\

Restoring Beam (diameter) & 5\farcs53\\

Total Integrated Time per Source & 35 minutes\\

RMS Noise in Q and U Maps (mJy/beam) & 0.029\tnote{b}\\

RMS Noise in RM Synthesis Maps ($\mu$Jy/beam) & 7.0\tnote{c} \\

\hline
\end{tabular}
\begin{tablenotes}
\item[a] The observations had 1.024 GHz wide intermediate frequency bands (IFs) centered on the frequencies listed.
\item [b] This number represents the average RMS noise level for all the Q and U maps.
\item[c] Polarized sensitivity of the combined RM Synthesis maps.
\end{tablenotes}
\end{threeparttable}
\end{table}

\begin{table}[p]
\centering
\caption{List of Sources Observed \label{tab:sources}}
\begin{threeparttable}
\centering
\begin{tabular}{ccccccc}
\hline
Source & $\alpha$(J2000) & $\delta$(J2000) & $\emph{l}$  & $\emph{b}$ & $\xi$\tnote{a} & I\tnote{b} \\
Name & h m s & $^o$ $'$ $''$  &  ($^o$) & ($^o$) & (arcmin) & (mJy beam$^{-1}$)  \\
\hline
I9 & 06 32 08.15 & 04 14 26.8 & 206.95  & -2.35 & 42.3 & 11.4  \\
I11 & 06 32 47.83 & 04 38 22.3 & 206.68 & -2.02 & 43.8 & 11.0\\
I13 & 06 33 28.81 & 04 08 37.4 & 206.19 & -2.09  & 53.3 & 1.9 \\
I17 & 06 34 38.88 & 05 36 42.5 & 206.02 & -1.16  & 57.2 & 14.0\\
N4 & 06 31 45.53 & 05 23 49.9 & 205.88 &-1.90  & 27.4 & 18.0\\
N5 & 06 32 52.66 & 05 21 45.9 & 206.04 & -1.67 & 29.0 &  3.8\\
N6 & 06 30 32.34 & 04 56 11.0 & 206.15 & -2.38 & 20.6 &  ---\\
N7 & 06 30 36.62 & 04 55 28.2 & 206.17  & -2.37 & 19.6 &  7.6\\
N8 & 06 30 41.82 & 04 55 14.2 & 206.18 & -2.35 & 18.3 &  ---\\
N9 & 06 30 28.13 & 04 41 58.6 & 206.35  & -2.50 & 26.1 &  2.2\\
N10 & 06 33 39.67 & 05 09 40.9 & 206.31  & -1.59  & 29.2 & 5.0\\
\hline
\end{tabular}
\begin{tablenotes}
\item[a] Angular distance between the line of sight and a line of sight through the center of the nebula.
\item[b] Stokes I intensity at 4.3 GHz. Note, value is for the (a) component when lines of sight has more than one.
\end{tablenotes}
\end{threeparttable}
\end{table}

\section{Data Reduction}\label{dataredux}
The data were reduced and imaged using the NRAO Common Astronomy Software Applications (CASA). The data were reduced following a similar procedure as in \citet{Savage:2013}.\footnote{\footnotesize{For further reference on data reduction, see the NRAO Jansky VLA tutorial ``EVLA Continuum Tutorial 3C391'' (http$://$casaguides.nrao.edu$/$index.php$?$title$=$EVLA$_{-}$Continuum$_{-}$Tutorial$_{-}$3C391)}} The procedure we implemented is as follows:
\begin{enumerate}
\item We initially flagged data using systematic flagging procedures (e.g., ``Quack'') before visually inspecting the data to manually remove data corrupted by radio frequency interference (RFI). In the frequency ranges we observed, there was minimal RFI except in two subbands, 7378--7506 MHz and 7506--7634 MHz, which were flagged completely.
\item Calibration of the data included determining complex gains, instrumental polarization parameters, and the R--L phase difference. After applying the solutions to the data, we ran a CASA systematic flagging mode in FLAGMANAGER, ``Rflag''. Rflag performs a second pass through a calibrated data set, reading the data in chunks of time and accumulating statistics to identify outlying data points. It then implements thresholds to auto flag these outlying data points. After running Rflag, a new calibration of the data was performed.
\item The calibrated visibility data were then imaged using the CASA task CLEAN to produce images in  Stokes I, Q, U, and V. CLEAN is a Fourier transform task which forms the ``dirty map'' and ``dirty beam'' of the data, implements the CLEAN deconvolution algorithm, and then restores the image by convolving the CLEAN components with the restoring beam. The restoring beam was set to 5\farcs53 for all the sources, and natural weighting was used in order to obtain the highest signal-to-noise-ratio (SNR). The images of the Stokes parameters I, Q, U, and V were made in two ways. These two sets of images were the input data for the two methods of determining the RM, described further in Sections \ref{sec:trad} and \ref{sec:syn}.
\begin{enumerate}
\item In the first approach, a single image of each Stokes parameter was made in 128 MHz-wide subbands within each 1.024 GHz intermediate frequency band (IFs).
The actual spectrum utilized was sightly less than 128 MHz because of discarded edge channels as well as channels lost to RFI. Since the bandwidth of 128 MHz is non-negligible compared to the center frequency, particularly for the lower frequency subbands, we used the ``mfs'' (multi-frequency synthesis) mode in the task CLEAN. This algorithm takes into account the different (\textit{u, v}) tracks corresponding to different frequencies. The image resulting from mfs CLEAN corresponds to a frequency at the center of the 128 MHz-wide subband. These images were the inputs to the \chilam~ analysis (Section \ref{sec:trad}).
\item The second approach was to make I, Q, U, and V images for each 4 MHz-wide piece of spectrum in our observations using the mode ``channel'' in CASA, which averages two adjacent 2 MHz channels. The resulting set of maps were used as input to the RM synthesis analysis (Section \ref{sec:syn}).
\end{enumerate}
\item Finally, phase-only self-calibration was performed on all sources. For the sources with sufficient SNR (i.e., SNR $>$ 20 )\footnote{\footnotesize{NRAO Data Reduction Workshop 2012, http://www.aoc.nrao.edu/events/synthesis/2012/lectures.shtml}}, we performed two iterations of phase-only self-calibration. The sources with  SNR $<$ 20 did not improve with phase-only self-calibration.
\end{enumerate}

\subsection{Determination of Rotation Measures Using Two Techniques\label{sec:twot}}
For all 11 sources, we employ two methods to determine the RM for each source or source component. The first method (Section \ref{sec:trad}) was implemented in \citet{Savage:2013} and consists of a least-squares linear fit of \chilam. This is a technique that has traditionally been used to measure Faraday rotation from radio astronomical polarization measurements. The second method (Section \ref{sec:syn}) is Rotation Measure Synthesis (RM Synthesis) \citep{Brentjens:2005}. RM Synthesis exploits the large, multi-channel data sets generated by modern interferometers like the VLA and avoids some of the shortcomings of the \chilam~fit. The following sections detail the imaging process for the two methods.

\subsubsection{Rotation Measures from Least-Squares Fit of \chilam \label{sec:trad}}
\begin{figure}[!ht]
\centering
\subfloat[\label{fig:N4}]{
\includegraphics[width=0.4\textwidth]{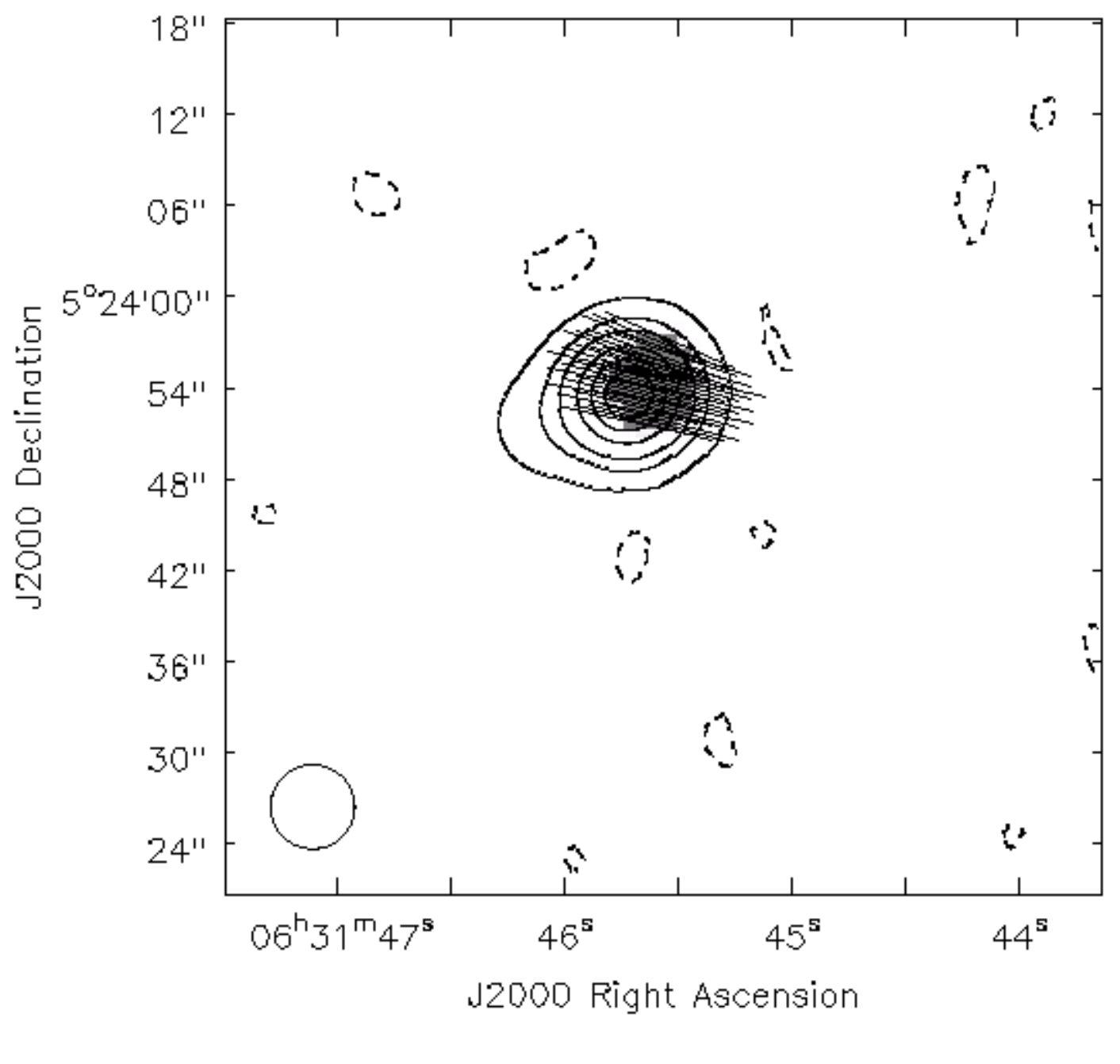}}
\quad
\subfloat[\label{fig:N10}]{
\includegraphics[width=0.42\textwidth]{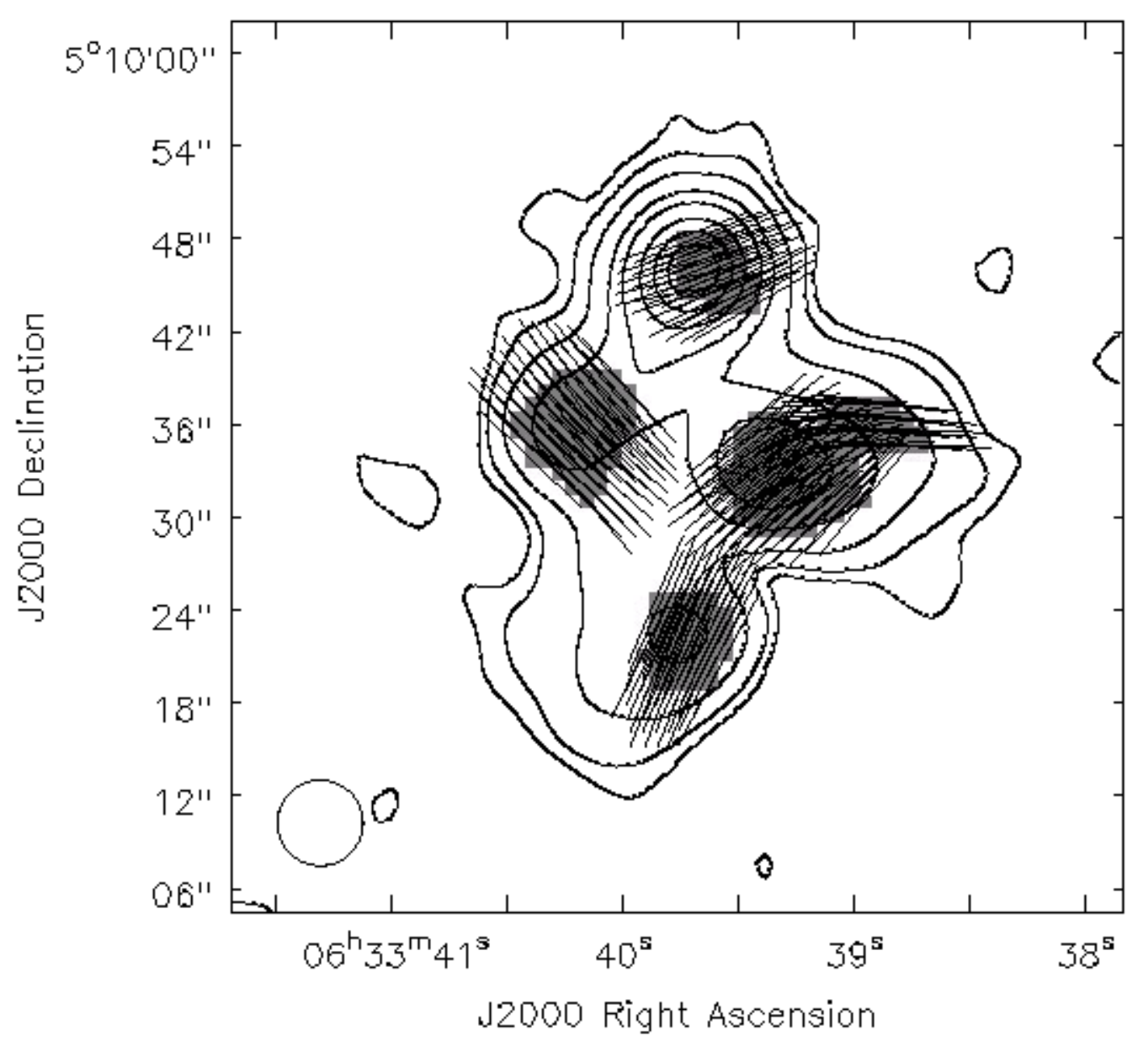}}
\caption[CASA Map]{Map of (a) N4 and (b) N10 at 4.85 GHz. The gray scale is the linear polarized intensity, P, the vectors show the polarization position angle, $\chi$, and the contours are the Stokes I intensity with levels of -2, -1 , 2, 10, 20, 40, 60, and 80$\%$ of the peak intensity, 24.5 mJy beam$^{-1}$ and 4.87 mJy beam$^{-1}$ for N4 and N10, respectively. The circle in the lower left is the restoring beam. In image (b), there are four resolved components of N10, and we call the northern component ``a'' in Table \ref{tab:results} and the following three components are marked ``b''--``d'' in a clockwise direction from ``a''.  }
\label{fig:MAP}
\end{figure}

The output of the CASA task CLEAN is a set of images in Stokes I, Q, U, and V. With these images, we use the CASA task IMMATH to generate maps of linear polarized intensity P, \[P=\sqrt{Q^2+U^2}, \] and the polarization position angle, $\chi$, \[\chi=\frac{1}{2}\tan^{-1}{\left(\frac{U}{Q}\right)}. \] Figures \ref{fig:N4} and \ref{fig:N10} are examples of maps for two sources, where the vectors are $\chi$, the gray scale is P, and the contours are the Stokes I intensity. Though some of the sources observed in this project are point sources to the VLA in C array, other sources like N10 (Figure \ref{fig:N10}) have resolvable structure. For each source, we produce maps in each of the the fourteen, 128 MHz-wide subbands. To obtain measurements of $\chi$, we select the pixel of the highest value of P of the source in the 4338 MHz map and measure $\chi$ at that location in each 128 MHz-wide subband.

\begin{figure}[hbt!]
\centering
\includegraphics[width=0.5\textwidth]{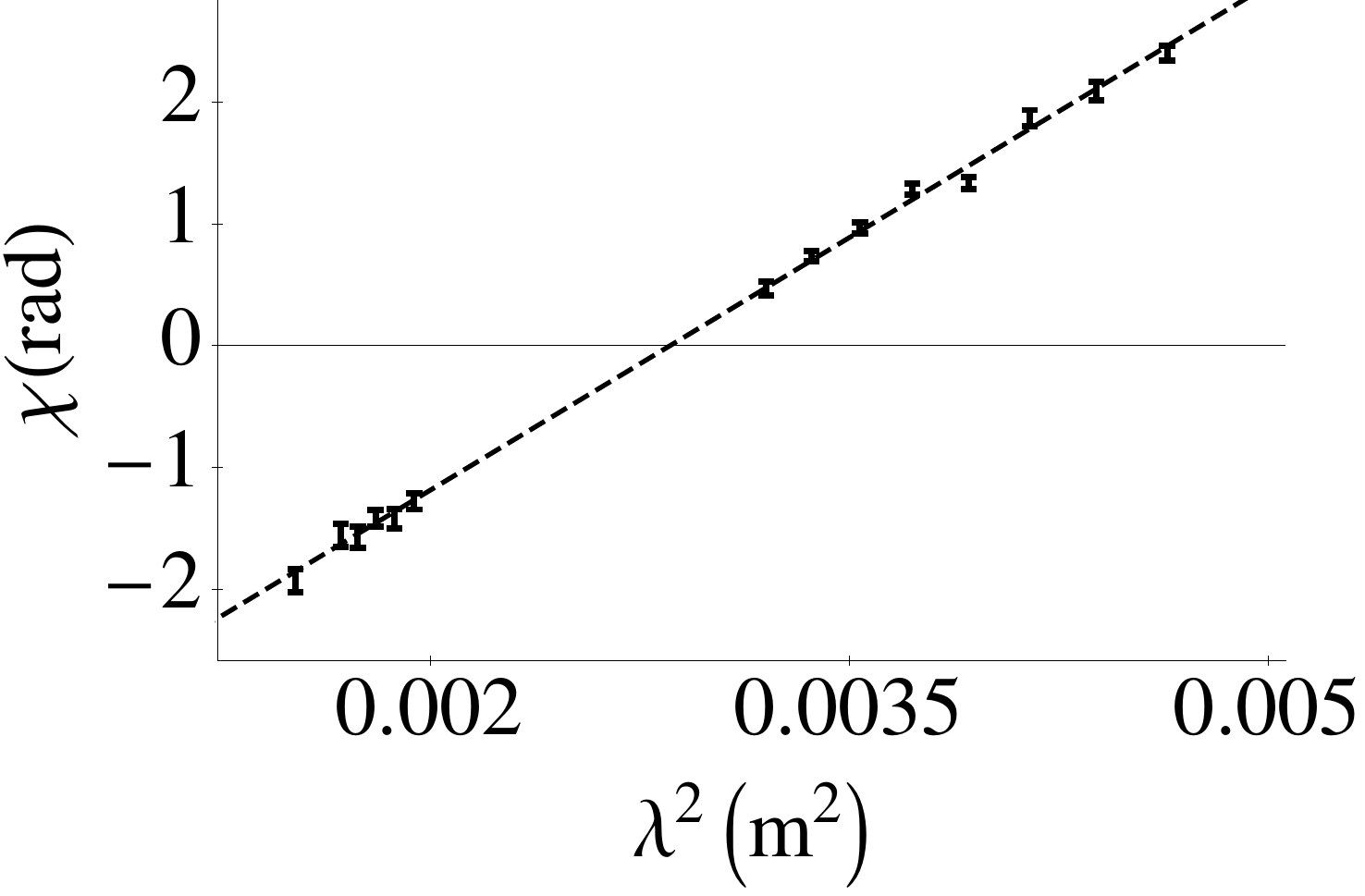}
\caption[Plot of $\chi(\lambda^{2})$ for N4]{Plot of the polarization position angle as a function of the square of the wavelength, $\chi(\lambda^{2})$, for the source N4, RM= +1383 $\pm$ 18 \radm. The reduced $\chi^{2}$ is 1.5. Each plotted point results from a measurement in a single 128 MHz-wide subband.}
\label{fig:newpol}
\end{figure}

\begin{figure}[!htb]
\centering
\subfloat[\label{fig:I9m}]{
\includegraphics[width=0.45\textwidth]{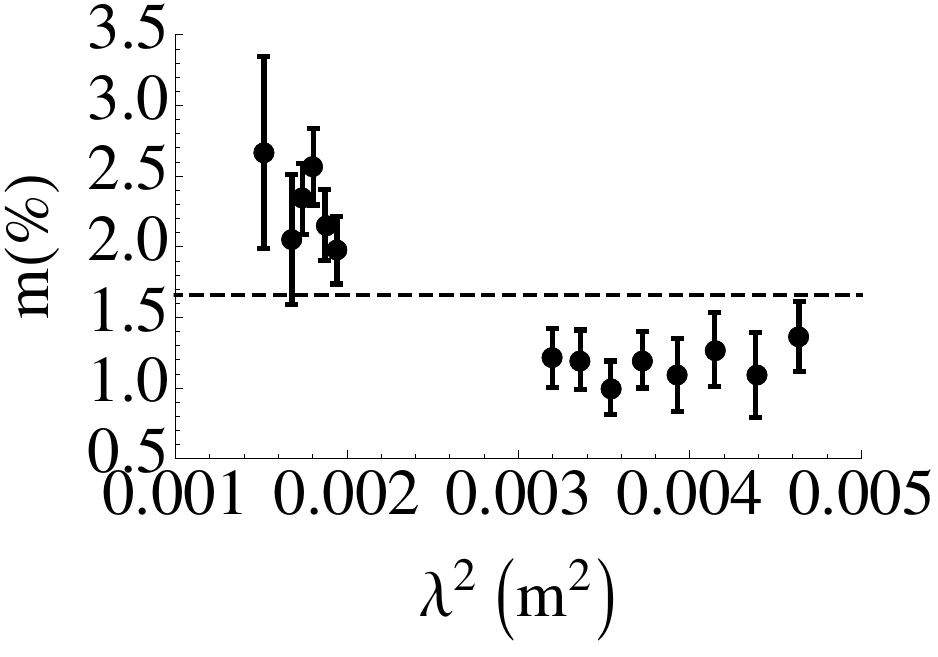}}
\quad
\subfloat[\label{fig:I17m}]{
\includegraphics[width=0.45\textwidth]{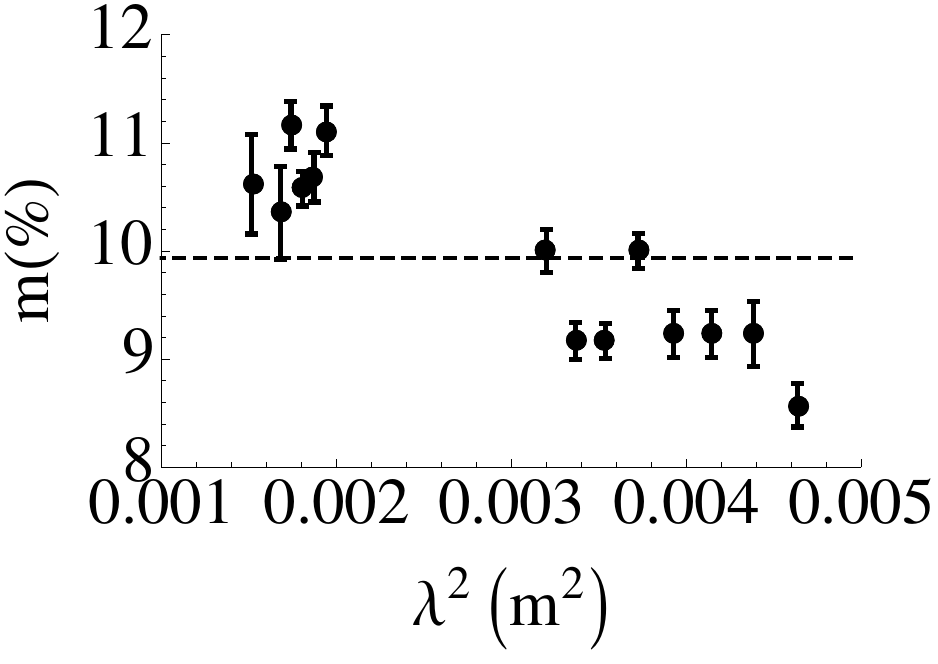}}
\caption[Percent Polarization vs Lambda Squared Graph]{Graph of the percent polarization, $m$, as a function of $\lambda^2$ for sources (a) I9 and (b) I17. These sources show a decrease in $m$ with increasing $\lambda^2$, which is characteristic of depolarization. The dashed line shows the median value of the percent polarization.}
\label{fig:mplot}
\end{figure}


We obtain RM measurements by carrying out a least-squares linear fit of \chilam, where RM is the slope of the line. 
Figure \ref{fig:newpol} shows an example of the least-squares fit for the source N4. The errors in $\chi$ are \(\sigma_{\chi}=\frac{\sigma_{Q}}{2P}~\)(\citealt{Everett:2001}, Equation (12)), where $\sigma_{Q}$ = $\sigma_{U}$ is the rms noise in the Q map. This method assumes that there is only one source along the line of sight, unaffected by beam depolarization and without internal Faraday rotation. Depolarization manifests itself as a change in m=P/I with frequency. 
Depolarization may occur for a number of reasons, such as external Faraday dispersion, multiple interfering RM components, differential Faraday rotation, or internal Faraday rotation \citep{Burn:1966,Sokoloff:1998,Osullivan:2012}. 

For all sources, we inspected plots of m($\lambda^2$) to determine if there were depolarization effects. Figures \ref{fig:I9m} and \ref{fig:I17m} show the graphs for the two sources, I9 and I17, respectively, which show depolarization. Despite the detection of depolarization for these two sources, we do not believe our resultant RM values to be biased or in error.  The reduced $\chi^2$ of the fit of \chilam~is satisfactory and suggests no departure from the $\chi$ vs $\lambda^2$ dependence (Table \ref{tab:results}). 

\subsubsection{RM Measurements via the Technique of RM Synthesis \label{sec:syn}}
The basic mathematics and physics of RM Synthesis was first discussed by \citet{Burn:1966}. However, the implementation of the fundamental idea with data from modern, wide bandwidth interferometers is due to \citet{Brentjens:2005}. With the upgraded continuous spectral coverage of 1--2 GHz of the VLA, it is possible to implement RM Synthesis for VLA polarimetric studies.


The Faraday dispersion function, F($\phi$), is a function of Faraday depth, $\phi$, and is related to the observed quantity, the complex polarized flux, $\tilde{P}$($\lambda^2$), where \(P = Q + \imath U\) \citep{Sokoloff:1998}. This function is defined as \(\tilde{P}(\lambda^2)=P(\lambda^2)W(\lambda^2), \) and W($\lambda^2$) is a weighting function that is zero for \(\lambda^{2}<0 \) \citep{Brentjens:2005,Heald:2009}. It is also zero for wavelengths at which observations do not exist, including wavelengths excised for RFI, and for wavelengths at which measurements exist, it is weighted by \(1/ \sigma^2 \), where $\sigma^2$ is the variance. Other possible weighting schemes include uniform weighting, where W($\lambda^2$) is unity for $\lambda^2>$0 and at the wavelengths at which measurements were made.

RM Synthesis utilizes a Fourier transform relationship to convolve F($\phi$) with the rotation measure spread function (RMSF). Mathematically, RMSF, or R$(\phi)$, is
\begin{equation}
R(\phi)=K\int_{-\infinity}^{\infinity}W(\lambda^{2}) e^{-2\imath\phi\lambda^{2}}d\lambda^{2}, 
\end{equation}
where $\lambda$ is the wavelength, and the output of the convolution is the reconstructed Faraday dispersion function, $\tilde{F}$($\phi$),
\begin{equation}
\tilde{F}(\phi)=F(\phi)\ast R(\phi)=K\int_{-\infinity}^{\infinity}\tilde{P}(\lambda^{2})e^{-2\imath\phi\lambda^{2}}d\lambda^{2},
\label{eq:rmsyn}
\end{equation}
 where $K$ is a normalization function given by \[K=\left(\int_{-\infinity}^{\infinity}W(\lambda^{2})d\lambda^{2}\right)^{-1}, \] (see \citet{Brentjens:2005} for the full derivation). To recover F($\phi$), $\tilde{F}$($\phi$) is deconvolved via a CLEAN algorithm such as the ones discussed in  \citet{Heald:2009} and \citet{Bell:2012}. In the case of a single point source behind a Faraday screen, F($\phi$) is a delta function at a Faraday depth equal to the RM through the screen. ``Faraday Complexity'' \citep{Anderson:2015} will lead to a broadening of the Faraday dispersion function, F($\phi$), and Q and U will have non-sinusoidal behavior. This broadening can also occur in the case of depolarization of a single component. 

\begin{figure}[htb!]
\centering
\includegraphics[width=0.5\textwidth]{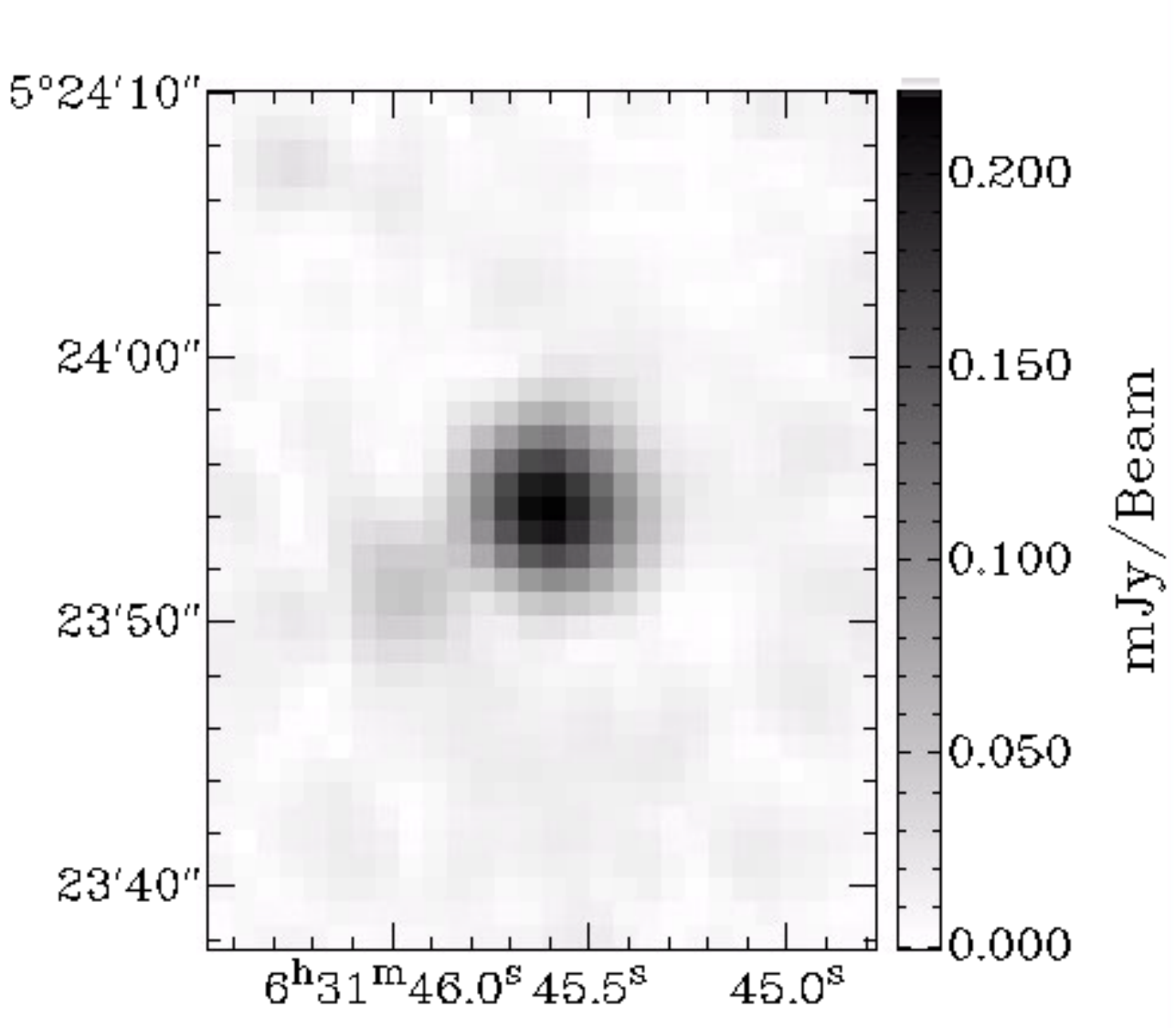}
\caption[RM Synthesis Map]{KVIS output map from the RM Synthesis analysis where the gray scale is the polarized intensity for the source N4 at $\phi$=1408 \radm.}
\label{fig:N4synmap}
\end{figure}

\begin{figure}[htb!]
\centering
\includegraphics[width=0.6\textwidth]{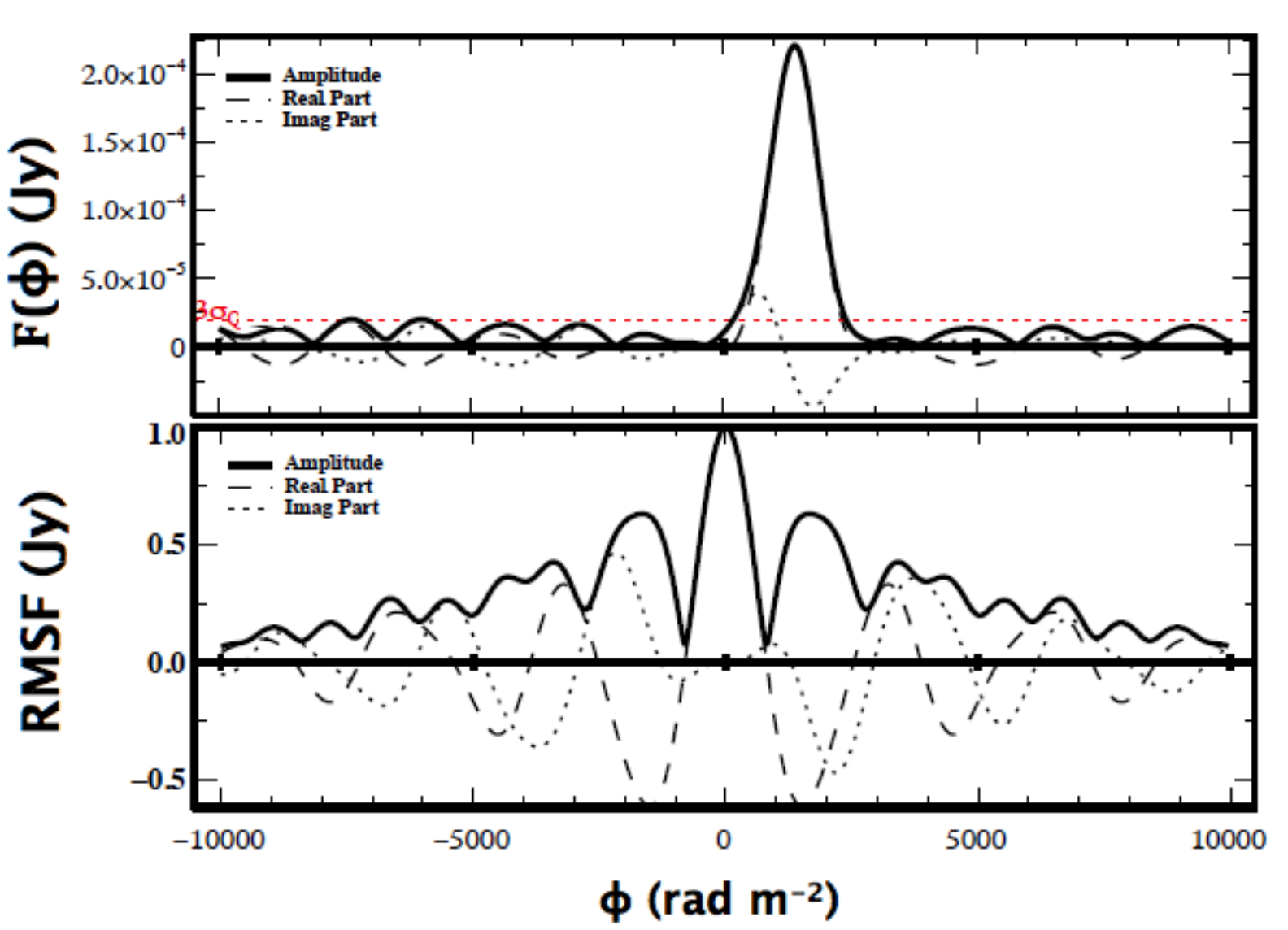}
\caption[RM Synthesis Data]{Top: Plot of the clean Faraday dispersion function, F($\phi$), for N4 at the position corresponding to the peak value of the linear polarized intensity determined in the \chilam~analysis. The curve peaks at +1408 $\pm$ 16 \radm.  The solid line is the amplitude, the dash-dot line is the real part, and the dashed line is the imaginary part. The red dotted line marks the 3$\sigma_{Q}$ threshold. Bottom: Plot of RMSF.}
\label{fig:N4syngraph}
\end{figure}

In practice, we utilize an IDL implementation of an RM Synthesis and CLEAN code. The inputs to the IDL code are FITS files of the CLEANed images of Stokes Q and U exported from CASA. The Q and U images are a function of frequency, Q($\nu$) and U($\nu$), and are composed of fourteen subbands containing twenty-four 4 MHz-wide channels\footnote{Ideally, an RFI free data set would permit the use of all sixteen 128 MHz-wide subbands containing thirty-two 4 MHz channels. In practice, the edge channels within each subband are typically flagged and thus not included in the final analysis.}. The data set input to the RM Synthesis code is therefore a set of 336 images in Q and 336 images in U. The frequency spacing between images is 4 MHz and is approximately the effective bandwidth in each image, except for gaps due to discarded edge channels between subbands. There are also gaps due to discarded or heavily flagged subbands. The output from the IDL code is an image in Faraday depth space, which retains the astrometric headers (e.g., R.A. and Dec.). As CASA cannot yet read images in Faraday depth space, we use the Karma package \citep{Gooch:1995} tool KVIS to read the images. In KVIS, we extract the polarized flux from a single pixel at the location of peak linear polarized intensity in each Faraday depth plane to acquire the Faraday dispersion function spectrum. We then fit a Gaussian to the Faraday dispersion function to recover the RM at F($\phi_{\textrm{{\scriptsize max}}}$). Multiple RM components may be present in the Faraday dispersion function and can be identified by comparing the data to the RMSF. The Faraday dispersion function was measured at the same spatial location as used for the \chilam~analysis, for each source or source component. This was done so that we can compare the two measurements.

For each source, we implemented a search range of  $\phi$ = $\pm$ 10000 \radm~to identify possible peaks at large values of $\phi$. We are sensitive up to $\phi_{\textrm{{\scriptsize max}}} \sim$ 3.0$\times$10$^{5}$ \radm, the full width at half-maximum (FWHM) of the RMSF is $\phi_{\textrm{{\scriptsize FWHM}}}$ = 1067 \radm, and the largest detectable scale in Faraday depth space (max-scale) is 2085 \radm (see Equations (61)--(63) in \citealt{Brentjens:2005}). In Table \ref{tab:results}, column 5 lists the effective RM values derived from the peak of the Faraday dispersion function from the RM Synthesis analysis.

Figure \ref{fig:N4synmap} shows an example of a KVIS map for source N4 at $\phi$ = 1408 \radm. The top panel of Figure \ref{fig:N4syngraph} is the Faraday dispersion function corresponding to one pixel from Figure \ref{fig:N4synmap}, and the bottom panel plots the RMSF. 
Using the spatial location of the peak linear polarized intensity from the \chilam~analysis, we select the same spatial location on the RM Synthesis map so that we may compare the two measurements. In general, this method samples the Faraday dispersion function at F($\phi_{\textrm{{\scriptsize peak}}}$) and at P$_{\textrm{{\scriptsize max}}}$ in the RM Synthesis data. The exception to this is N9(b). The spatial location of F($\phi_{\textrm{{\scriptsize max}}}$) in the RM Synthesis map does not coincide with the location in the \chilam~maps, which is most likely due to the values of P not exceeding the 5$\sigma_Q$ threshold that is implemented in the imaging process in CASA (see Section \ref{sec:comp}) in the \chilam~ analysis. The reported RM value derived from the RM Synthesis analysis is spatially coincident with the \chilam~value to maintain consistency.

\section{Observational Results \label{sec:res} }
\begin{figure}[htb!]
\centering
\includegraphics[width=0.5\textwidth]{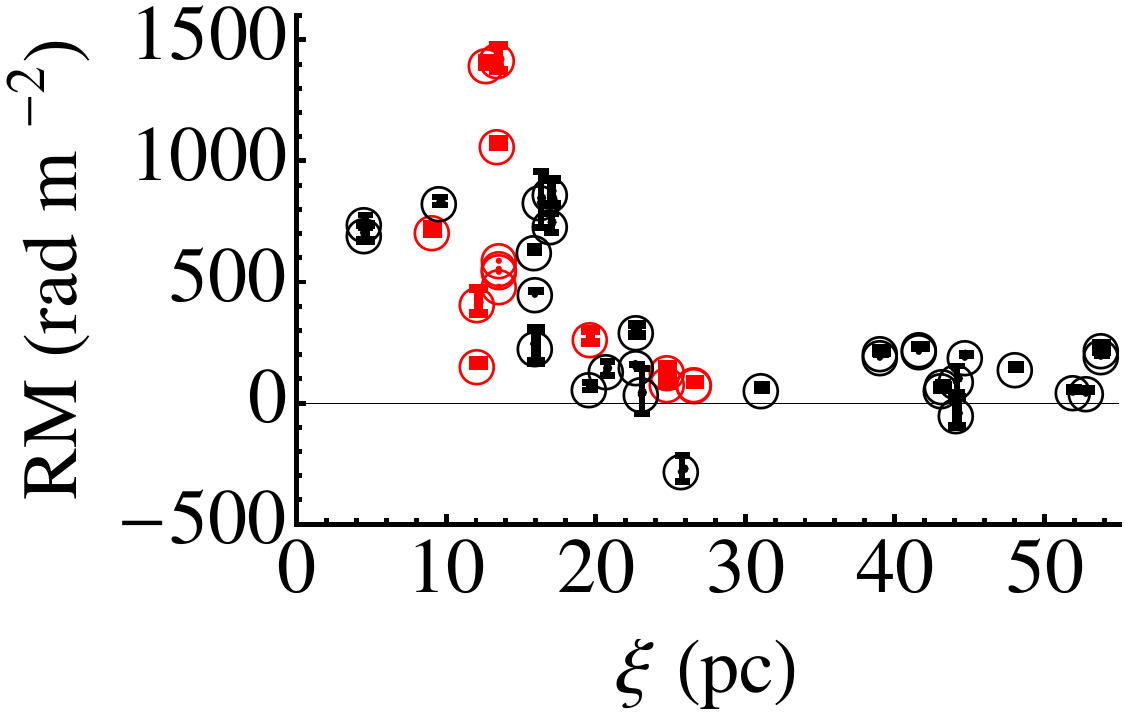}
\caption[Observed RM as a Function of Distance]{Plot of observed RM as a function of distance in parsecs from the center of the Rosette Nebula. The lines of sight observed in this paper are denoted by red circles, and the plotted RM values are derived from the peak of the Faraday dispersion function. The results from \citet{Savage:2013} are the black circles. All of the sources have error bars; in many cases, the size of the measurement error is smaller than the plotted point. Secondary source components are represented.}
\label{fig:distlos}
\end{figure}

For the 11 sources observed through the shell of the Rosette Nebula, we measured RM values for fifteen lines of sight, including secondary components. Table \ref{tab:results} lists the RM values and associated errors for the least-squares method and from the RM Synthesis analysis.  We have three lines of sight that do not pass through the shell of the nebula, I9, I13, and I17. Including the secondary components, we measure an average background RM due to the general ISM of +146 $\pm$ 37 \radm. The background RM is in perfect agreement with the value from \citet{Savage:2013}.  We see no evidence for a gradient in the measured background RM values over the 4\ddeg~diameter region centered on the nebula. We measure an excess RM ranging from +40 to +1200 \radm~due to the shell of the nebula.  Figure \ref{fig:rosettenewlos} shows the new sources in red in combination with the results from \citet{Savage:2013}, where the symbols are scaled to \(\sqrt{|RM|}. \) Figure \ref{fig:distlos} plots the observed RM measurements as a function of distance from in the center of the Rosette Nebula in parsecs, using a distance of 1600 pc to the nebula \citep{Roman:2008}. 

\begin{table}[hbt!]
\centering
\begin{threeparttable}
\caption{New Faraday Rotation Measurements through the Rosette Nebula \label{tab:results}}
\begin{tabular}{ccccc}
\hline
\multirow{2}{*}{Source} & \multirow{2}{*}{Component} &  RM\tnote{a} & Reduced  & RM\tnote{c} \\
 &  & (\radm) & $\chi^{2}$\tnote{b} & (\radm) \\
\hline 
\multirow{1}{*}{I9} & a    & +318$\pm$24 & 1.2 &  +276$\pm$26 \\
\hline 
\multirow{1}{*}{I11} & a  &  ---& ---& ---\\ \hline 
\multirow{2}{*}{I13} & a & +39$\pm$33  & 3.2 & +90$\pm$15 \\
 & b  &  +130$\pm$24 & 0.6 & +139$\pm$19 \\ \hline 
\multirow{2}{*}{I17} & a  & +81 $\pm$4  & 2.1 & +83$\pm$3 \\ 
& b& +116$\pm$8 & 0.8 & +87$\pm$10 \\ \hline 
\multirow{1}{*}{N4} & a  & +1383$\pm$18  & 1.5 &+1408$\pm$16 \\ \hline
\multirow{2}{*}{N5} & a  & +1062$\pm$21 & 1.8 & +1074$\pm$14 \\
 & b  &   +1332$\pm$74 & 2.1 & +1426$\pm$52 \\ \hline 
\multirow{1}{*}{N6} & a  & --- & ---&  ---\\ \hline 
\multirow{1}{*}{N7} & a   & +697$\pm$17 & 0.8 & +719$\pm$19 \\ \hline 
\multirow{1}{*}{N8} & a  & --- & --- &  ---\\ \hline 
\multirow{2}{*}{N9} & a   & +175$\pm$20 & 1.9 & +165$\pm$9 \\ 
 & b &   +571$\pm$98 & 1.3 & +421$\pm$51 \\ \hline 
\multirow{4}{*}{N10} & a   & +619$\pm$42  & 1.3 & +569$\pm$29\\
 & b  &   +614$\pm$49 & 8.4 & +556$\pm$14 \\
 & c  &   +507$\pm$20 & 1.4& +495$\pm$23 \\
 & d  &   +664$\pm$40 & 9.2 & +603$\pm$26\\ \hline 
\end{tabular}
\begin{tablenotes}
\item[a] RM value obtained from a least-squares linear fit to $\chi(\lambda^2)$.
\item[b] Reduced $\chi^{2}$ for the \chilam~fit.
\item[c] Effective RM derived from RM Synthesis
\end{tablenotes}
\end{threeparttable}
\end{table}

Double or multiple sources with large changes in RM values can probe small scale structure in the Rosette Nebula. We use the variable $\Delta$RM to indicate differences in the RM between components or parts of a source with angular separations larger than the restoring beam of 5\farcs5. Of the 8 sources with RMs, 5 sources have more than one component that yield RM values. N10 has a maximum angular separation of 53$''$ between the northern and the southern component and 13$''$ between the eastern and western components. For I13 and I17, the angular separations are 30$''$ and 10$''$, respectively. In these cases, the $\Delta$RM values are relatively small, nearly within the errors in the case of I17. The $\Delta$RM values for N5 and N9, however, appear to be significant, with $\Delta$RM $\sim$ 350 and 260 \radm~for N5 and N9, respectively. The angular separation between the components for N5 is 18$''$, and N9 has a separation of 50$''$. The linear separations between the components for these sources are 0.14 and 0.39 pc, respectively. Our measurements of $\Delta$ RM may indicate small scale gradients in the electron density or line-of-sight component of the magnetic field in the Rosette Nebula.  These variations could be due to inhomogeneities in the shell, or turbulent fluctuations in $n_e$ and $B_{\parallel}$ on spatial scales of 0.1 to 1 parsec.

We did not obtain RM measurements for three of the sources, I11, N6, and N8. The sources N6, N7, and N8 were closely-spaced sources taken from the NVSS survey, with a maximum angular separation of 1.4 arcmin. We imposed a cutoff in the uv plane, which excluded data with uv distances $<$ 5000 wavelengths in the CLEAN task for these sources. For the VLA in C array at C-band, this (u, v) cutoff should eliminate any foreground emission due to the nebula and not emission from the extragalactic sources. N6 and N8 were completely filtered out by this process. Source I11 is very weakly polarized or unpolarized at 4.3--7.7 GHz and did not improve with phase-only self-calibration. The RM synthesis technique allows us to determine that I11 is unpolarized (no significant peak in the F($\phi$) plot), and we can exclude the possibility that it possesses an extremely large RM that would result in depolarization over the 128 MHz-wide subband.

\subsection{Comparison of Techniques for RM Measurement \label{sec:comp}}
\begin{figure}[htb!]
\centering
\includegraphics[width=.6\textwidth]{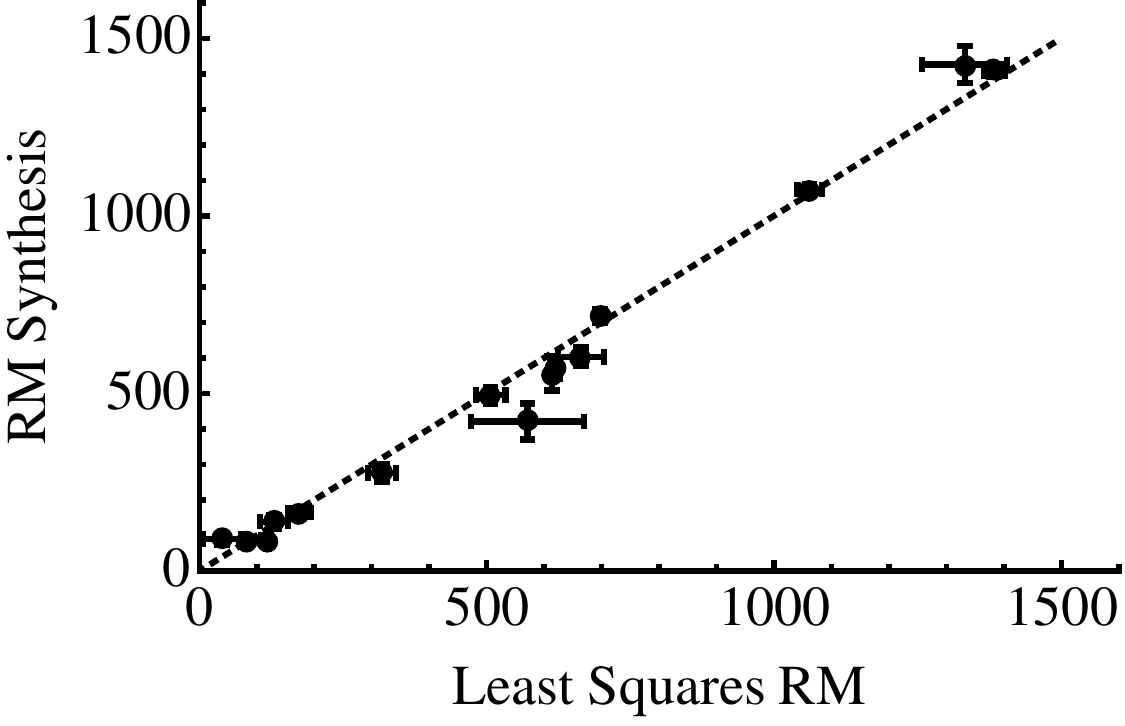}
\caption[Comparison of RM Synthesis to \chilam~Technique]{Plot comparing the RM measurements (\radm) obtained from RM Synthesis and the traditional least-squares fitting method of \chilam. The line represents perfect agreement between the two measurements, and errors for both methods are represented.}
\label{fig:comparison}
\end{figure}

In this section, we discuss the results of the traditional \chilam~method of obtaining RM values and RM Synthesis. For each of the sources and source components in Table \ref{tab:results}, we have two measurements of RM, one from each of the two techniques. Figure \ref{fig:comparison} plots the $\chi(\lambda^2)$ RM values against the RM measurements obtained through RM Synthesis. The line shows the case of perfect agreement between the two methods. There is very good agreement between the two measurements. The RM Synthesis technique has higher precision, evident in the smaller errors, since it uses the entire 2 GHz bandwidth to calculate the RM. We feel the results of Figure \ref{fig:comparison} lend confidence to the RM values we report. 

The only source that deviates strongly is component (b) of N9.
Component (b) of N9 is a weak source. The Stokes I intensity at 4.3GHz $\sim$1.2 mJy beam$^{-1}$. N9(b) did not improve with phase-only self-calibration, and the linear polarized intensity only exceeds the 5$\sigma_{Q}$ threshold in one subband in the \chilam~analysis. N9(b) is present, however, in the RM Synthesis analysis. RM Synthesis has the advantage over the \chilam~method of being less sensitive to low SNR levels than the traditional \chilam~method.  In spite of the large errors associated with the components of N9, the difference in $\Delta$RM is large enough to be significant. In our analysis in the following sections, we use the RM values derived from the RM Synthesis technique for all of the sources.

\section{Diagnostics of Magnetic Fields in \HII Regions Utilizing Plasma Shell Models \label{section:shells} }

\subsection{Empirical Models for Stellar Bubbles\label{sec:emmodels}}
\citet{Savage:2013} employed a physically motivated shell model to represent the magnitude and spatial distribution of the observed RM in the shell of the Rosette Nebula. 
The equation describing this shell model is
\begin{equation}
RM(\xi)=\frac{Cn_{e}L(\xi)}{2}[B_{zI}+B_{zE}],
\end{equation}
where $B_{zI}$ and $B_{zE}$ are the line of sight components of the magnetic field in the shell at ingress and egress, respectively, and $C$ is a set of constants equal to 0.81 defined by Equation (\ref{RMSI}) when L($\xi)$, \textbf{B}, and n$_{e}$ are measured in parsecs, $\mu$Gauss, and cm$^{-3}$, respectively (see Equation (10) in \citealt{Whiting:2009}). B$_{zI}$ and B$_{zE}$ are defined to be in the shell, and downstream from the hypothesized outer shock that defines the outer limit of the bubble. Relations between B$_{zI}$ and B$_{zE}$ and the upstream, undisturbed ISM magnetic field are given in Equations (6)--(9) of \citet{Whiting:2009}. L($\xi$) is the chord length through the shell, and it has two domains given by
\begin{equation}
L(\xi)=2R_{0}\sqrt{\left(1-\left(\frac{\xi}{R_{0}}\right)^{2}\right)} \text{, if } \xi \geq R_{1} \text{ , and }\\
\label{eq:lengths}
\end{equation}
\begin{equation*}
L(\xi)=2R_{0}\left[\sqrt{\left(1-\left(\frac{\xi}{R_{0}}\right)^2\right)}-\left(\frac{R_{1}}{R_{0}}\right)\sqrt{\left(1-\left(\frac{\xi}{R_{1}}\right)^2\right)}\right]\text{,} \text{ if } \xi \leq R_{1} \text{,} 
\end{equation*}
where R$_{0}$ and R$_{1}$ are the outer and inner radii of the shell, respectively, and $\xi$ is the distance between the line of sight and the center of the shell (see Figure 6 of \citealt{Whiting:2009}). Outside the outer radius of the \HII region (R$_{0}$ $<$ $\xi$), the model predicts RM=0, so the background RM due to the Galactic plane in this region is determined empirically through observations in the vicinity of, but outside the Rosette Nebula. The external general interstellar magnetic field, \textbf{B}$_{0}$, can be treated as a constant, meaning we assume \textbf{B}$_{0}$ does not vary significantly over the extent of the Rosette Nebula, i.e., outside the nebula, \textbf{B}$_0$ is constant in magnitude and direction. At the shock front located at R$_{0}$, \textbf{B}$_{0}$ can be decomposed into components perpendicular (B$_{\perp}$) and parallel (B$_{\parallel}$) to the shock normal. The perpendicular component is amplified by the density compression ratio (X) in the shell. Use of all these considerations then gives (\citealt{Savage:2013}, Equation (8))
\begin{equation}
RM=Cn_{e}L(\xi)B_{0z}\left(1+(X-1)\left(\frac{\xi}{R_{0}}\right)^{2}\right).
\label{RM3}
\end{equation}
B$_{0z}$ is the z-component of \textbf{B}$_{0}$, such that
\begin{equation}
B_{0z}=B_{0}\cos{\theta},
\label{eq:B}
\end{equation}
where B$_0$ is the magnitude of the ISM magnetic field and $\theta$, the angle between the direction to Earth and \textbf{B}$_{0}$, becomes the only free parameter in the model. We assume B$_{0}$ to be a constant and known. K. Ferri\`{e}re (private communication) has pointed out that in reality, there will be fluctuations in $|$B$_0$$|$ as well as $\theta$ in the interstellar medium. As a result, it can be argued that B$_{0z}$ should be considered the true independent parameter in Equation (\ref{RM3}). In the limit of a purely adiabatic strong shock, i.e., X=4, Equation (\ref{RM3}) predicts that the maximum value of the RM occurs near the outer radius of the shell for $\xi$ $<$ $R_{0}$. As discussed in Section \ref{sec:prev}, \citet{Harvey:2011} consider a case in which the increase in RM is due solely to an increase in density and not the magnetic field. In this limit where X=1, Equation ({\ref{RM3}}) models the expected RM in the shell without an enhancement in the magnetic field,
\begin{equation}
RM(\xi)=Cn_{e}L(\xi)B_{0z}
\label{RM2}
\end{equation}
and it predicts the maximum value of the RM is at the contact discontinuity, R$_{1}$.

\begin{figure}[!htb]
\centering
\subfloat[\label{fig:model1}]{
\includegraphics[width=0.45\textwidth]{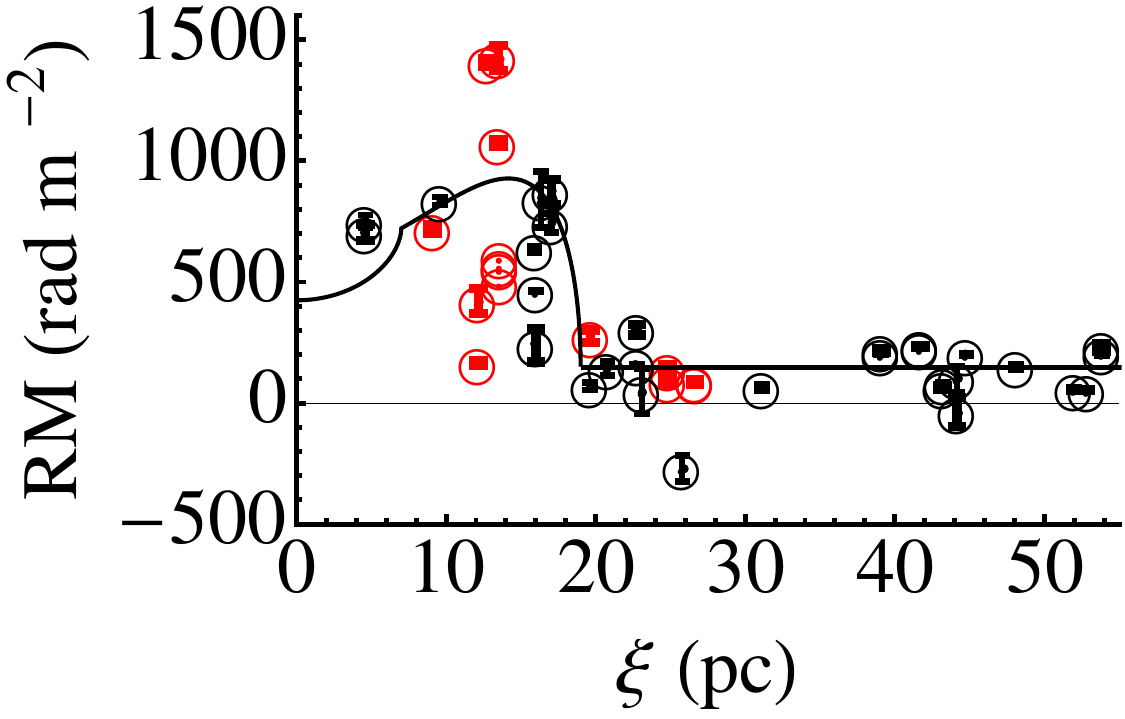}}
\quad
\subfloat[\label{fig:model2}]{
\includegraphics[width=0.45\textwidth]{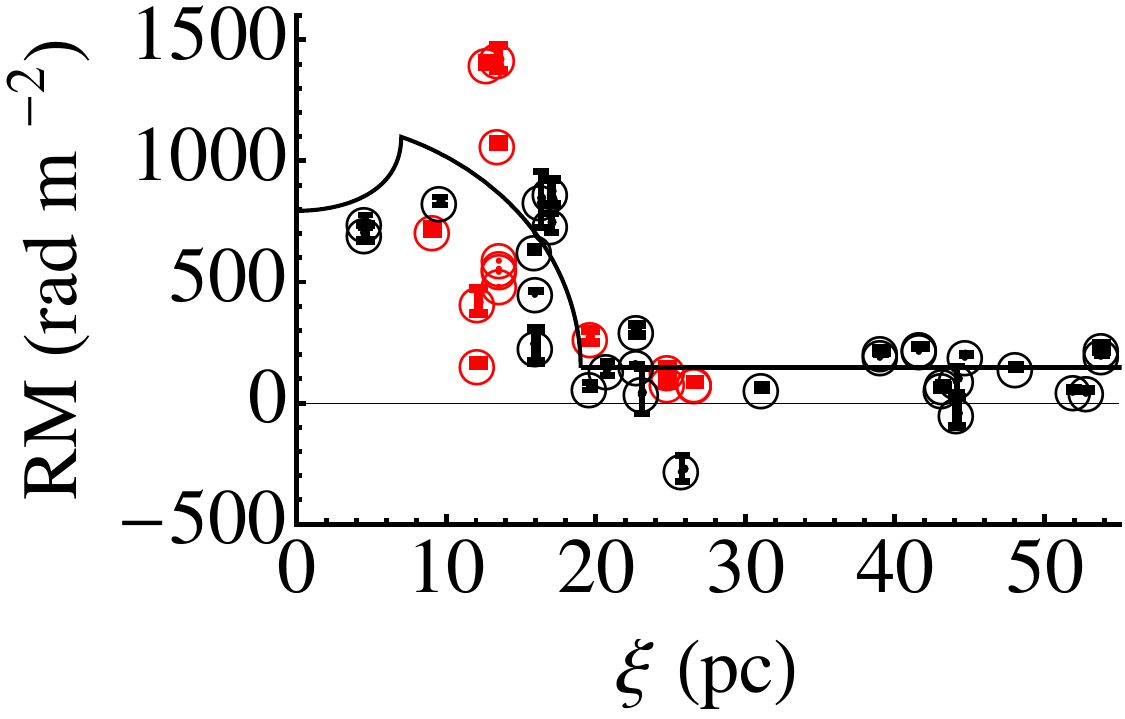}}
\caption[Shell Model]{Graph similar to Figure \ref{fig:distlos} with models superimposed for (a) Equation (\ref{RM3}) with $\theta$ = 73\ddeg and (b) Equation (\ref{RM2}) with $\theta$ = 47\ddeg.}
\label{fig:2model}
\end{figure}

Equations (\ref{RM3}) and (\ref{RM2}) represent two different models for the structure of the bubble, albeit highly simplified and approximate. The model described by Equation (\ref{RM3}) posits an increase in the interstellar magnetic field in the shell as well as an increase in the plasma density that is incorporated in the empirical parameter n$_{e}$. A comparison between the models of Equations (\ref{RM3}) and (\ref{RM2}) and the data is shown in Figure \ref{fig:2model}. Figure \ref{fig:model1} represents the model with shock-enhanced magnetic field (Equation (\ref{RM3})), and Figure \ref{fig:model2} is the model with no modification of the interstellar field (Equation (\ref{RM2})). Both models incorporate shell parameters (R$_0$, R$_1$, and n$_e$) from Celnik's Model I \citep{Celnik:1985}.

The models plotted here differ slightly from those discussed in \citet{Savage:2013} . In the present paper and \citet{Savage:2013}, we adopt a distance for the Rosette Nebula of 1600 parsecs, whereas \citet{Celnik:1985} assumed a distance of 1420 pc. In the present paper, we have therefore scaled up R$_0$ and R$_1$ by 13$\%$ and correspondingly lowered the value of n$_e$. The new values are given in Table \ref{tab:newpar}. This correction was noted by \citet{Planck:2015}. Given the rough approximation of the Celnik models to the structure of the Rosette Nebula, the model used in \citet{Savage:2013} is still useful as one approximation of the nebular structure. The values of $\theta$ in Table \ref{tab:newpar} and used in Figure \ref{fig:2model} are slightly modified from \citet{Savage:2013}. These new values of $\theta$ are the result of the Bayesian analysis discussed in Section \ref{sec:bay}.

\begin{table}[hbt!]
\centering
\begin{threeparttable}

\caption{Input Parameters for Shell Models \label{tab:newpar}}
\begin{tabular}{cccccc}
\hline
 R$_0$ & R$_1$ & n$_e$ & B$_0$ & X\tnote{a} & $\theta$\tnote{b} \\
(pc) & (pc) & (cm$^{-3}$) & ($\mu$G) &  & (deg) \\ \hline
19.0 & 7.0 & 12.2 & 4 & 4 & 73 \\
 19.0 & 7.0& 12.2 &4 & 1 & 47 \\ \hline
\end{tabular}
\begin{tablenotes}
\item[a] A value of 4 is for a purely adiabatic shock, and 1 is for no modification of magnetic field.
\item[b] Values of $\theta$ derived from the Bayesian analysis in Section \ref{sec:bay}.
\end{tablenotes}
\end{threeparttable}
\end{table}
\subsubsection{Physical Context of Bubble Models}
The physical content of the two models discussed here is described in Section \ref{sec:emmodels} above, as well as \citet{Savage:2013} and \citet{Whiting:2009}.  Both are subject to criticisms as regards physical content, for reasons we briefly consider here.

The analysis of \citet{Harvey:2011} may be incomplete because it is difficult to see how the general interstellar magnetic field would not be modified by advection and compression due to the highly conducting plasma shell of the nebula.  However, the model of \citet{Whiting:2009} is subject to criticism for reasons brought to our attention by K. Ferri\`{e}re (private communication).  The main physical ingredient of the \citet{Whiting:2009} expression is the modification of the interstellar field \textbf{B}$_0$ by the strong outer shock wave assumed to be present at R$_0$.  Given the ionized or partially ionized state of this part of the interstellar medium, this will be an MHD shock.  The \citet{Whiting:2009} model applies the magnetic field jump conditions that occur at an MHD shock (e.g., \citet{Gurnett:2005}).

Aside from the question as to whether a strong shock exists in this part of the Rosette Nebula, the point raised by Dr. Ferri\`{e}re addresses the extent of the amplified field throughout the shell.  In the \citet{Whiting:2009} expression, it is assumed that the amplified field applies to a large portion of the shell (see fuller discussion in \citealt{Whiting:2009}). Dr. Ferri\`{e}re's point is that while this simplification might be valid in the case of a thin shell, it becomes questionable in the case of an object like the Rosette Nebula, in which the shell between R$_1$ and R$_0$ occupies most of the volume of the nebula.

The points raised by Dr. Ferri\`{e}re are included in \citet{Planck:2015} (Section 5 and Appendix A).  That paper also includes a model of the nebular field based on the assumption of conservation of magnetic flux.  While recognizing the points raised above, we will continue to use the expressions contained in Equations (\ref{RM3}) and (\ref{RM2}) as part of the discussion in the present paper.  These expressions represent limiting behavior of the magnetic field within a bubble shell.  For example, Equation (\ref{RM3}) may be considered a proxy for any mechanism that would enhance the interstellar field within the nebular bubble, and rotate it into a plane perpendicular to the radial direction.  It is perhaps worth pointing out that the RM($\xi$) curves presented by \citet{Planck:2015} in their Figure 5 also show ``rotation measure limb brightening'' that is characteristic of the Whiting model. In Figure 5(a) of \citet{Planck:2015}, the peak of the curve is located at $\xi\sim$12 -- 13 pc for the blue and red curves, respectively. This peak is only slightly interior to where the peak of Equation (\ref{RM3}) is located at $\xi\sim$ 14.2 pc. In both Equations (\ref{RM3}) and (\ref{RM2}), the discontinuity in the derivative of RM at R$_1$ results from the assumption that a contact discontinuity occurs here.  Furthermore, both models assume the region interior to R$_1$ is a vacuum. In any case, the comparison of measurements to a set of simplified descriptions of magnetic field behavior should further the goal of these investigations, which is to better specify the magnetic field within these shells, and how they are related to the general interstellar magnetic field.

\subsection{A Bayesian Statistical Approach to Compare Shell Models and Determine the Orientation of the Interstellar Magnetic Field\label{sec:bay}}
In this section, we carry out a Bayesian statistical analysis to determine which of the models described in Section 5.1 better reproduces the observed RM measurements, and therefore offers a better representation of the magnetic field in the Rosette Nebula.  The Bayesian analysis is suitable for our purposes because the two models may be considered ``hypotheses'' that are being tested \citep{Gregory:2005}. Our approach and terminology follows that presented in \citet{Gregory:2005}.  We adopt from \citet{Celnik:1985} the parameters for the shell such as the inner and outer radius, R$_{1}$ and R$_{0}$, and the electron density, n$_{e}$. These parameters have been scaled to the distance to the Rosette Nebula of 1600 pc. As in \citet{Planck:2015}, we adopt R$_{1}$ = 7.0 pc, R$_{0}$ = 19.0 pc, and n$_e$ = 12.2 cm$^{-3}$. We retain the value of 4 $\mu$G for the magnitude of the external magnetic field, B$_{0}$ \citep{Ferriere:2011}. Given the measured and adopted values of the parameters, the RM through the bubble is determined by the angle $\theta$. In \citet{Savage:2013}, a value of $\theta$ = 72\ddeg~for a shell with an amplified magnetic field was obtained by visually inspecting the fit of the model to the data. For the model without the amplification of the magnetic field (Equation (\ref{RM2})), $\theta$ = 54\ddeg. The Bayesian analysis to be presented below automatically yields the optimum value of $\theta$ for each model. Although the primary reason for carrying out the Bayesian analysis is to determine which model better represents the observations, irrespective of the value of $\theta$, this quantity is nonetheless of astronomical interest.

Bayes' Theorem states that the probability of a hypothesis being true given some statement of prior information, is \[p(M_{i}|D,I)=\frac{p(M_{i}|I)p(D|M_{i},I)}{p(D|I)}, \] where M$_{i}$ is a model we are testing, I is a statement representing our prior information, and D is the data set. The denominator \(p(D|I) \) is referred to as the ``global likelihood'' \citep{Gregory:2005}. The likelihood function is
\begin{equation}
p(D|M,I)\equiv \int_{0}^{\pi/2} d\theta  \mbox{ } p(\theta|M,I)p(D|M,\theta,I),
\end{equation}
and p($\theta$$|$M,I) is the prior. Each hypothesis is a function of the unknown ``nuisance parameter'', $\theta$, which must be marginalized to compare two different hypothesis (shell models in our case) \citep{Gregory:2005}. The value of $\theta$ can range from 0 to \(\frac{\pi}{2} \) on the surface on a hemisphere. From this treatment, we will know the likelihood function, regardless of the value of $\theta$. We can, however, obtain estimates of $\theta$ as well. Our choice of p($\theta$$|$M,I) allows $\theta$ to range on the surface of a hemisphere, \[p(\theta|M,I)=\sin{\theta}. \]
The data consist of $N$ elements $d_i$, where $N$ is the number of RM measurements (19 in the present case of $\xi$ $\leq$ R$_0$). We may write d$_i$=RM$_i$+e$_i$, where RM$_i$ is the RM predicted by the model (Equation (\ref{RM3}) or Equation (\ref{RM2})), and e$_i$ is simply the difference between the model and the measured value. The probability of a model being true can be represented as a product of the independent probabilities for each measurement, and if the e$_i$ are independent (see Section 4.8 in \citealt{Gregory:2005}), then
\begin{equation}
p(D|M,\theta,I) = \prod_{i}^{N}p(d_{i}|M,\theta,I)=\prod_{i}^{N}\frac{1}{\sigma\sqrt{2\pi}}\exp{\left(-\frac{e_{i}^{2}}{2\sigma^{2}}\right)},
\label{eq:bay1}
\end{equation}
where $\sigma^2$ is the variance. The likelihood function for a model M is then
\begin{equation}
p(D|M,I)=(2\pi)^{-N/2}\int_{0}^{\pi/2}d\theta \mbox{ }\sin{\theta} \mbox{ }\\ \prod_i^N\sigma_i^{-N}\mbox{ } \exp{\left(\frac{-(d_{i}-RM_{i})^{2}}{2\sigma_i^{2}}\right)} \mbox{. }
\label{eq:pD}
\end{equation}
In practice, we perform an explicit summation over $\theta$ to obtain the likelihood functions, p(D$\mid$M$_1$,I) and p(D$\mid$M$_2$,I).

We can compare the two models by computing the \textit{odds ratio} \citep{Gregory:2005}, \[O_{12}=\frac{p(M_{1}|D,I)}{p(M_{2}|D,I)}. \] Assuming the prior probabilities are equal, p(M$_1|$I)=p(M$_2|$I), the odds ratio reduces to the Bayes factor and is then a ratio of the likelihood functions
\begin{equation}
B_{12}=\frac{p(D|M_{1},I)}{p(D|M_{2},I) }\mbox{.}
\end{equation}
To select the preferred model, we use the `Jeffreys' scale' (Table 1 in \citealt{Trotta:2008}), which categorizes the evidence as ``inconclusive'', ``weak'', ``moderate'', or ``strong'' based on the value of the Bayes factor. \citet{Gordon:2007} use the variable $B$ to denote the Bayes factor such that on the Jeffreys' scale, \(|\ln{B}| < 1 \) is inconclusive, \(1 \leq|\ln{B}| \leq 2.5 \) is weak, \(2.5 \leq |\ln{B}| \leq 5 \) is moderate, and \(5 \leq |\ln{B}| \)  is strong evidence.

We adopt the following prescription for $\sigma$ in the Bayesian analysis. In Equation (\ref{eq:bay1}), the values of $\sigma_{i}$ recognize the fact that the two models described by Equations (\ref{RM3}) and (\ref{RM2}), while physically motivated, are approximations at best.  With this viewpoint, the measurement errors on RM, determined by radiometer noise, are irrelevant. The main factor governing the departure of the measurements from the model is the crudeness of the model. We allow the model to deviate from the observed RMs by a scale factor, $\alpha$, such that \(\sigma_i=\alpha RM_{i},\) where RM$_{i}$ is the calculated model value of the RM. There is no rigorous way of specifying the parameter $\alpha$. We determine $\alpha$ as follows:
\begin{enumerate}
\item We perform the Bayesian analysis with an arbitrary value of $\alpha$ = 0.2 to determine \thetaP, which is the value of $\theta$ at maximum likelihood, $\mathscr{L}_{\textrm{max}}$, where  \[\mathscr{L}_{\textrm{max}} = \Bigl[p(D|M,\theta,I)\Bigr]_{\textrm{max}}\textrm{,} \] for both models.
\item Using the values of \thetaP, we calculate RM values for each line of sight with Equations (\ref{RM3}) and (\ref{RM2}) and find the distribution of \[x_i=\frac{d_i-\textrm{RM}_i}{\textrm{RM}_i}. \] 
\item We then calculate the standard deviation, $\sigma_{\textrm{SD}}$, of the distribution of $x_i$, and we then perform another iteration of the Bayesian analysis with $\alpha$ = $\sigma_{\textrm{SD}}$. From our analysis, we determined that $\alpha$ equals 0.41 and 0.43 for the homogeneous and inhomogeneous (to be defined in Section \ref{sec:joe} below)  models, respectively.
\end{enumerate}

\begin{figure}[htb!]
\centering
\subfloat[\label{fig:exp1}]{
\includegraphics[width=0.4\textwidth]{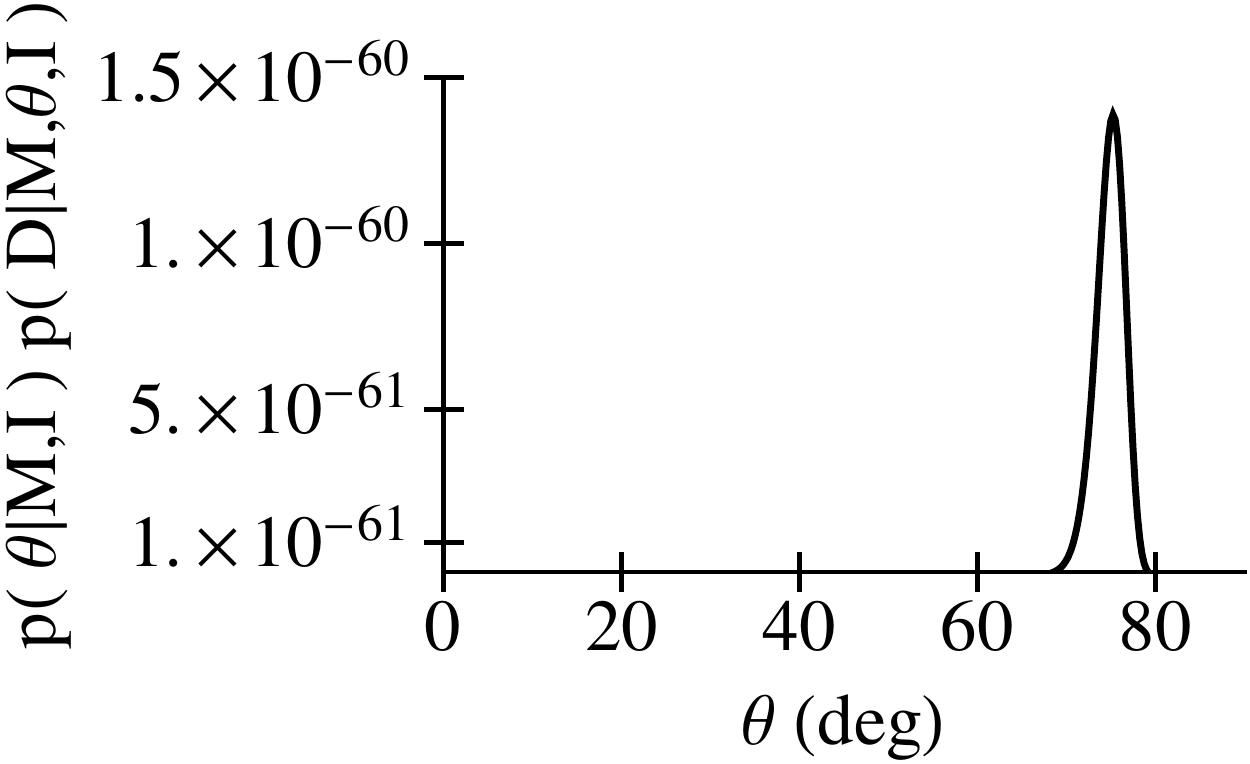}}\quad
\subfloat[\label{fig:exp2}]{
\includegraphics[width=0.4\textwidth]{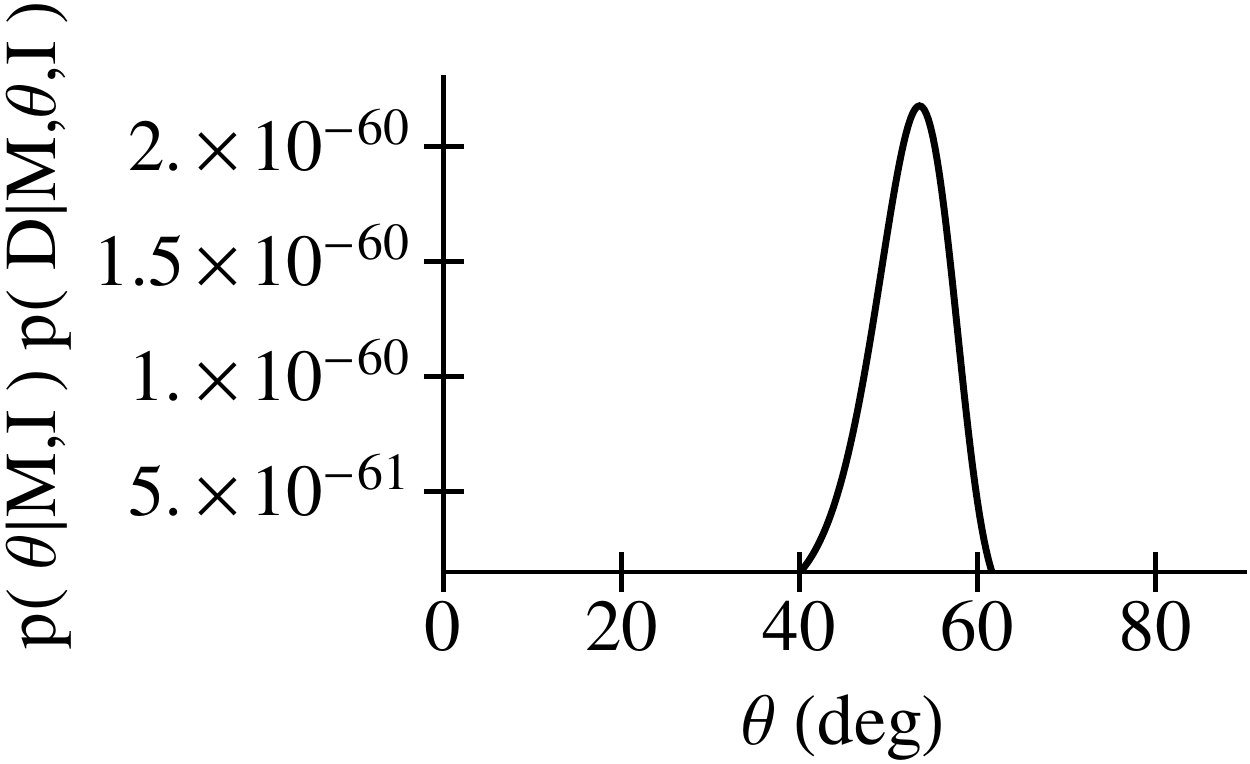}}\\
\subfloat[\label{fig:local1}]{
\includegraphics[width=0.4\textwidth]{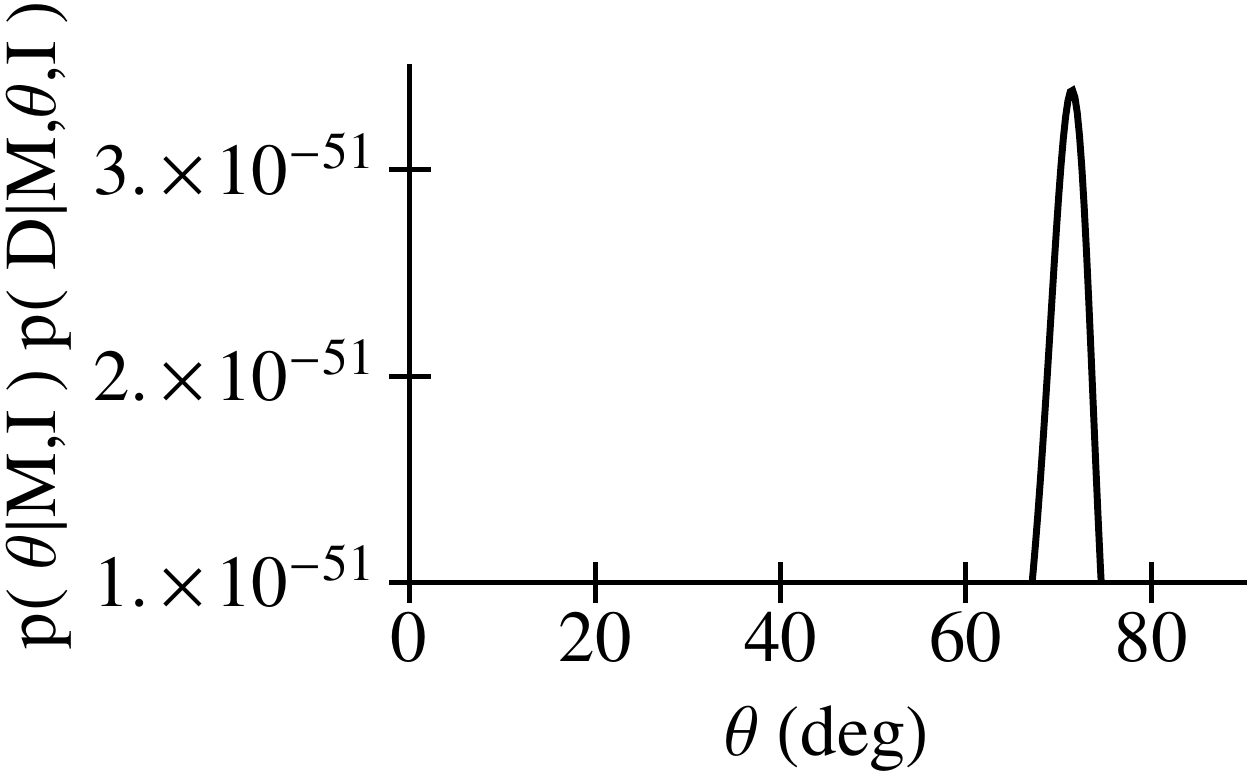}}\quad
\subfloat[\label{fig:local2}]{
\includegraphics[width=0.4\textwidth]{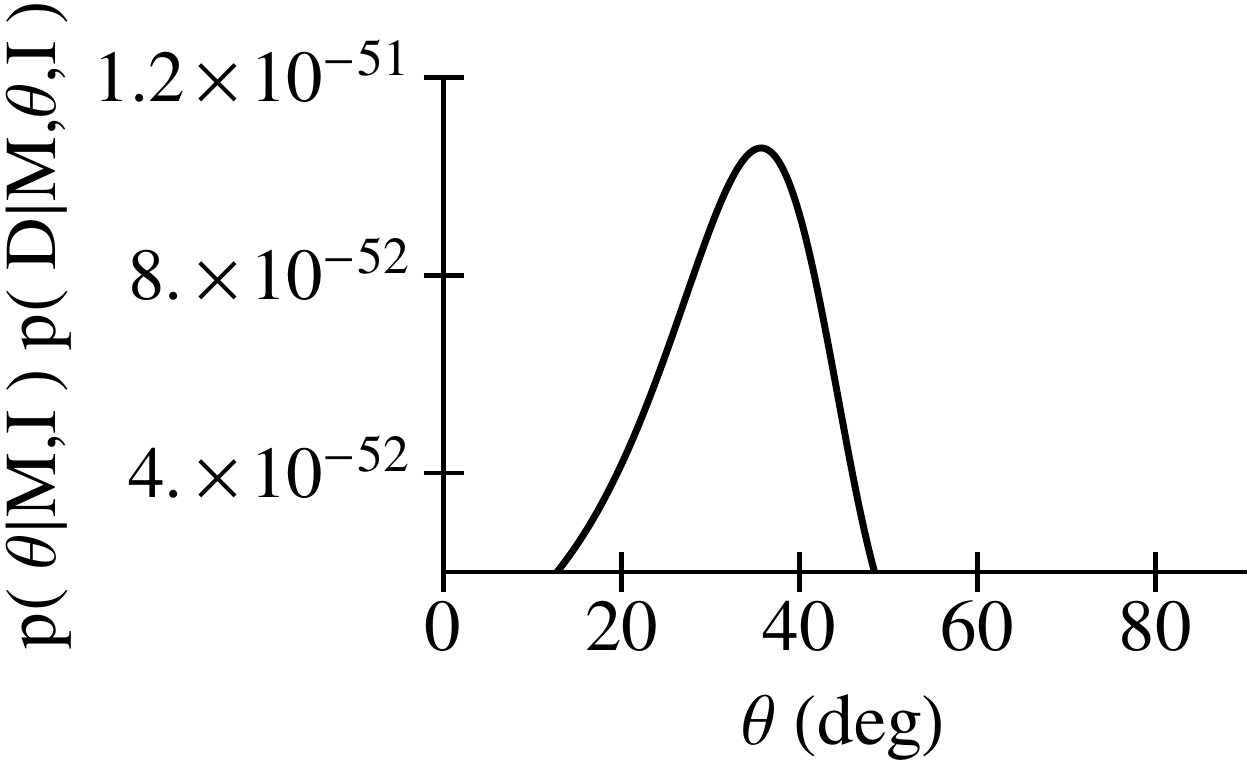}}

\caption[Plots of Likelihood functions as a function of $\theta$]{Plots of p($\theta$$\mid$M,I)p(D$\mid$M,$\theta$,I) from Equation (\ref{eq:pD}) as a function of $\theta$ in increments of $d\theta$= 0.25\ddeg. The values of the shell model parameters are listed in Table \ref{tab:newpar} for (a) the shell model with an amplification of the magnetic field in the shell (Equation (\ref{RM3})) and (b) the shell model without an amplification of the magnetic field in the shell (Equation (\ref{RM2})). The values for the  inhomogeneous shell model parameters are listed in Table \ref{tab:localpars} (see Section \ref{sec:joe}) for (c) the model with an amplification of the magnetic field and (d) the model without an amplification.}
\end{figure}

The results of the Bayesian analysis are listed in Table \ref{tab:alpha}. The first column lists the type of model, homogeneous or inhomogeneous, used in the Bayesian analysis. Column two lists the value of $\alpha$, and column three gives the value of density compression ratio, X. The likelihood function, p(D$\mid$M$_i$,I), is reported in column four, and the corresponding value of \thetaP~is in column five. Columns six and seven give B$_{12}$ and (B$_{12})^{-1}$, respectively. Finally, column eight lists $|\ln{B}|$ and the corresponding strength of evidence on the Jeffreys' scale is given in column nine. 
Figures \ref{fig:exp1} and \ref{fig:exp2} plot the integrand of Equation (\ref{eq:pD}) for the models with fixed values of  R$_1$, R$_0$, and n$_e$ and where $\alpha$ = 0.41. For the model with an amplification of the magnetic field in the shell (Equation (\ref{RM3})), p(D$\mid$M$_1$,I) = 4.1$\times$10$^{-54}$ with \thetaP~= 75\ddeg. For the model without an amplification (Equation (\ref{RM2})), p(D$\mid$M$_2$,I) = 2.2$\times$10$^{-53}$ with \thetaP~= 54\ddeg. The Bayes factor is B$_{12}$ = 0.19 or B$_{12}^{-1}$ = 5.4. Thus, \(|\ln{B}| = 1.7\), and this is considered ``weak'' evidence in favor of the model without an amplification of the magnetic field in the shell, i.e., that considered by \citet{Harvey:2011}, on the Jeffreys' scale. In this analysis, we have restricted the shell radii and the electron density to be global parameters for the shell. The Bayesian analysis does not conclusively favor either model.  There is a weak tendency to favor the model without modification of the general interstellar field.



\begin{table}[ht!]
\centering
\begin{threeparttable}

\caption{Results of Bayesian Analysis\label{tab:alpha}}

    \begin{tabular}{ccccccccc}
\hline
Shell  & $\alpha$  & X &  p(D$|$M$_i$,I)  & \thetaP (deg) & B$_{12}$ & (B$_{12}$)$^{-1}$ & $| \ln{B} |$ & Jeffreys' Scale \\ \hline
\multirow{2}{*}{Homogeneous\tnote{a}} & \multirow{2}{*}{0.20} & 1 & 1.9 $\times$ 10$^{-69}$ & 47 & \multirow{2}{*}{0.06} &  \multirow{2}{*}{16.0} & \multirow{2}{*}{2.8} & \multirow{2}{*}{moderate}\\
 & & 4 & 9.9 $\times$ 10$^{-71}$ & 73 & & & & \\
\multirow{2}{*}{Homogeneous\tnote{a}} & \multirow{2}{*}{0.41} & 1 & 2.2 $\times$ 10$^{-53}$ & 54 & \multirow{2}{*}{0.19} &  \multirow{2}{*}{5.4} & \multirow{2}{*}{1.7} &  \multirow{2}{*}{weak} \\
 & & 4 & 4.1 $\times$ 10$^{-54}$ & 75 & & & & \\
\multirow{2}{*}{Inhomogeneous\tnote{b}} & \multirow{2}{*}{0.20} & 1 & 5.4 $\times$ 10$^{-61}$ & 17 & \multirow{2}{*}{11} &  \multirow{2}{*}{0.1} & \multirow{2}{*}{2.4} & \multirow{2}{*}{weak}\\
 & & 4 & 6.1 $\times$ 10$^{-60}$ & 68 & & & & \\
\multirow{2}{*}{Inhomogeneous\tnote{b}} & \multirow{2}{*}{0.43} & 1 & 2.6 $\times$ 10$^{-45}$ & 33 & \multirow{2}{*}{0.37} &  \multirow{2}{*}{2.7} & \multirow{2}{*}{1.0} & \multirow{2}{*}{weak}\\
 & & 4 & 9.2 $\times$ 10$^{-46}$ & 71 & & & & \\

\hline
\end{tabular}
\begin{tablenotes}
\item[a] {\footnotesize For global shell values of R$_0$ = 19.0 pc, R$_1$ = 7.0 pc, and n$_e$ = 12.2 cm$^{-3}$}
\item[b] {\footnotesize Localized shell parameters as described in Section \ref{sec:joe}.}


\end{tablenotes}
\end{threeparttable}
\end{table}

\subsection{Simplified Inhomogeneous Shell Models\label{sec:joe}}

\begin{table}[p]
  \centering
  \begin{threeparttable}
  \caption{Local Parameters for Lines of Sight Through the Shell\label{tab:localpars}}
    \begin{tabular}{cccccccc}
\hline
    Source & T$_B$ & $\xi$\tnote{a} & R$_0$ & R$_1$ & n$_e$ & RM\tnote{b}    & RM\tnote{c} \\
          &   (K)    & (pc)    & (pc)    & (pc)    & (cm$^{-3}$) & (\radm)   & (\radm) \\
\hline
    I7\tnote{d,e}   &   0.49$\pm$0.13    & 4.60   & 15.1$\pm$0.3  & 9.2$\pm$0.1   & 17.6$\pm$3.0  &+454$\pm$73 & +764$\pm$176  \\
    I8\tnote{d, f}    &    0.46$\pm$0.04   & 16.0    & 19.1$\pm$0.4  & 8.40$\pm$.3   & 13.4$\pm$0.7  & +1063$\pm$59& +909$\pm$62 \\
    I10\tnote{d, g}   &   0.30$\pm$0.03    & 16.4  & 18.8$\pm$0.2  & 7.4$\pm$0.1   & 11.5$\pm$0.5  & +881$\pm$38 & +724$\pm$33  \\
    I12\tnote{d, h}   &   0.59$\pm$0.09    & 9.60   & 18.2$\pm$0.4  & 11.3$\pm$0.1  & 16.0$\pm$1.6    & +726$\pm$65 & +961$\pm$141\\
    N4    &   0.51$\pm$0.04    & 12.7  & 17.7$\pm$0.4  & 7.4$\pm$0.1   & 13.0$\pm$0.6    & +1005$\pm$41 & +1014$\pm$53 \\
    N5    &   0.44$\pm$0.05    & 13.5  & 16.5$\pm$0.5  & 7.4$\pm$0.1   & 13.7$\pm$1.0  & +973$\pm$75 & +855$\pm$82  \\
    N7    &   0.75$\pm$0.07    & 9.10  & 18.2$\pm$0.5  & 7.9$\pm$0.1   & 13.9$\pm$0.7  & +955$\pm$42 & +1338 $\pm$75\\
    N9    &  0.70$\pm$0.08     & 12.1  & 16.5$\pm$0.5  & 9.8$\pm$0.4   & 15.8$\pm$1.1  & +1127$\pm$72 & +1110$\pm$89 \\
    N10   &   0.12$\pm$0.05    & 13.6  & 14.1$\pm$0.4  & 5.2$\pm$0.1   & 11.4$\pm$3.4  & +500$\pm$162 & +388$\pm$120 \\    \hline
\end{tabular}
\begin{tablenotes}
\item[a] Distance between the line of sight and the line of sight through the center of the nebula at a distance of 1.6kpc.
\item[b] RM value calculated by Equation (\ref{RM3}) using $\theta$ =71\ddeg and $\alpha$ = 0.43.
\item[c] RM value calculated by Equation (\ref{RM2}) using $\theta$ = 33\ddeg and  $\alpha$ = 0.43.
\item[d] Source from \citet{Savage:2013}
\item[e] Component (a) RM = +749 $\pm$ 27 \radm~and component (b) RM = +704 $\pm$ 33 \radm.
\item[f] Component (a) RM = +461 $\pm$ 4 \radm~and component (b) RM = +238 $\pm$ 73 \radm.
\item[g] RM = +841 $\pm$ 117 \radm.
\item[h] RM = +835 $\pm$ 17 \radm.
\end{tablenotes}
\end{threeparttable}
\end{table}
In \citet{Savage:2013}, we assumed that the shell is spherically symmetric and the density is homogeneous as described by Model I of \citet{Celnik:1985}. However, the real structure of the Rosette Nebula has variation in both the inner and outer radii R$_1$ and R$_0$ (and thus the shell thickness (R$_0$-R$_1$)) as well as the density n$_e$. The simple shell model implemented in Section \ref{section:shells} is heavily dependent on these three parameters. In Equation (\ref{eq:lengths}), the location of the contact discontinuity (R$_1$) determines whether the line of sight resides in the region of shocked stellar wind closest to the star (\(\xi<R_1 \)), within the shell of shocked ISM gas (\(R_1 \leq \xi \leq R_0)\), or upstream of the shock in the ambient ISM (\(\xi>R_0\)).

In reality, the Rosette Nebula is not a spherically-symmetric object, although it is a better approximation to this ideal than most \HII regions. The lack of spherical symmetry, visible on optical or radio images, is doubtlessly due to the inhomogeneous nature of the medium into which the bubble expands. As a result, the local effective values of R$_0$, R$_1$, and n$_e$ will differ from one source line of sight to another. This would obviously affect our analysis of which shell model better represents the observed RM.

\begin{figure}[htb!]
\centering
\includegraphics[width=0.55\textwidth]{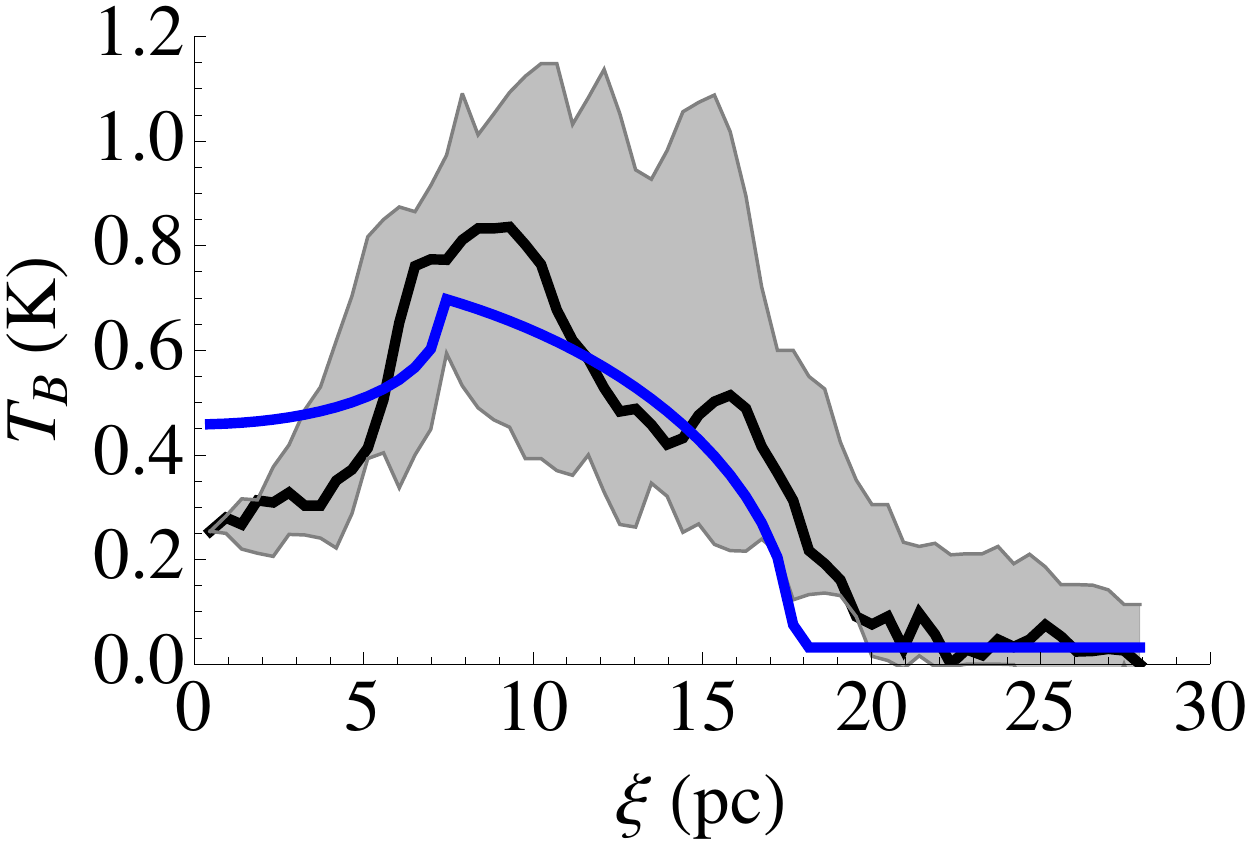}
\caption[Fit to Radio Continuum Data for N4]{Plot of the measured brightness temperature from the 4.95 GHz radio continuum map as a function of $\xi$ from the center of the Rosette Nebula in parsecs to the source N4. The black line is the data, the blue line is the fit to the data, and the shaded gray region represents the range in measured T$_B$ for the sources in Table \ref{tab:localpars}. The fitted parameters are the inner and outer radii and the electron density.}
\label{fig:shellfit}
\end{figure}

In this section, we discuss our method of obtaining the local values for the inner and outer shell radii and the electron density. By doing so, we can marginally relax the assumptions of spherical symmetry and homogeneous density within the shell. In essence, we are fitting a homogeneous spherical shell to a slice of the thermal radio emission passing through each source. The inhomogeneity of the Rosette Nebula is then accommodated by allowing these parameters to vary for different slices through the nebula. The result is that we obtain a set of R$_0$, R$_1$, and n$_e$ for each source. Consequently, we are using the brightness temperature profile through each source to the nebular edge to estimate the line-of-sight path length.

To determine the local shell parameters, we used the archival 4.95 GHz radio continuum map of the Rosette Nebula from \citet{Celnik:1985}, made at the Max Planck Institut f\"ur Radioastronomie, and generously furnished to us by Dr. Wolfgang Reich. We measured the brightness temperature, T$_{B}$, as a function of $\xi$ along slices through the nebula at different position angles $\psi$. The data consist of measurements of brightness temperature that we then convert to density by first calculating the emission measure, EM,
\begin{equation}
\mbox{EM}=282\mbox{ T}_{8}^{0.35}\mbox{ }\nu_{\mbox{\footnotesize{GHz}}}^{2.1}\mbox{ T}_{T} \mbox{ cm}^{-6}\mbox{ pc}
\label{eq:tb}
\end{equation}
(Equation (19) of \citet{Spangler:1998}), where $T_{8}$ is the electron temperature in units of 8000 K, $\nu_{\mbox{\footnotesize{GHz}}}$ is the frequency of the observation in units of GHz, and $T_{T}$ is the thermal component of the brightness temperature in units K. In the case of the Rosette Nebula, we make the assumption that all of the radio emission above the background level is thermal, T$_T$=T$_B$. The emission measure is related to the electron density by
\begin{equation}
\mbox{EM}= \int n_{e}^2 dz=fn_{e}^2  L, 
\label{eq:ne}
\end{equation}
where $n_{e}$ is the electron density in cm$^{-3}$, $f$ is the filling factor, and $L$ is the length of the path through the shell in pc, given by Equation (\ref{eq:lengths}). For present analysis we assume a filling factor $f = 1$; further discussion of the role of the filling factor is given in Section \ref{sec:filling}. We then carry out a least-squares fit of the model to the observed brightness temperature along the slice. To determine shell parameters n$_{e}$, R$_{0}$, and R$_{1}$ for a given slice, we use Equations (\ref{eq:tb}) and (\ref{eq:ne}) and Equation (\ref{eq:lengths}) for the chord length L($\xi$). For example, Figure \ref{fig:shellfit} plots the measured brightness temperature as a function of $\xi$ for one source, N4, as well as the range in brightness temperature profiles for the interior sources listed in Table \ref{tab:localpars}. We fit for the inner and outer radii, and then calculate the electron density for each line of sight.

Since we obtained estimates of the electron density and shell radii as a function of the angle $\psi$, we calculate the expected RM on a specific line of sight using the local values for the radii and electron density and with values of $\theta$ determined from the Bayesian analysis (see Section \ref{sec:bayjoe}.) The results of this analysis are listed in Table \ref{tab:localpars} for sources with $\xi \leq $ R$_0$ from both this work and \citet{Savage:2013}. A disadvantage of this approach is that three source components, I6a, I6b, and I15 were excluded from the inhomogeneous shell model analysis.  In these three cases, the source was beyond the  model outer extent of the shell, defined by the fit value of R$_0$.  As such, a meaningful model estimate for RM was not available.  Examination of the 4.95 GHz Effelsburg map showed that there was indeed plasma along the line of sight to each source, as indicated by thermal radio emission, but that brighter emission closer to the nebular center dominated the fit for the local shell model.  These sources do not therefore pose any problem in our understanding, but this situation did mean that the number of source/source components was reduced from 19 in the case of the homogeneous shell analysis to 16 for the inhomogeneous shell model.

The first column of Table \ref{tab:localpars} lists the source defining the line of sight, and column 2 lists the measured brightness temperature at the position of the source. The value of $\xi$ for the source is listed in column 3. Columns 4, 5, and 6 list the estimates of R$_0$, R$_1$, and n$_e$, respectively. Finally, columns 7 and 8 give the calculated RM values from Equations (\ref{RM3}) and (\ref{RM2}), respectively.

\begin{figure}[htb!]
\centering
\subfloat[\label{fig:SRM}]{
\includegraphics[width=0.4\textwidth]{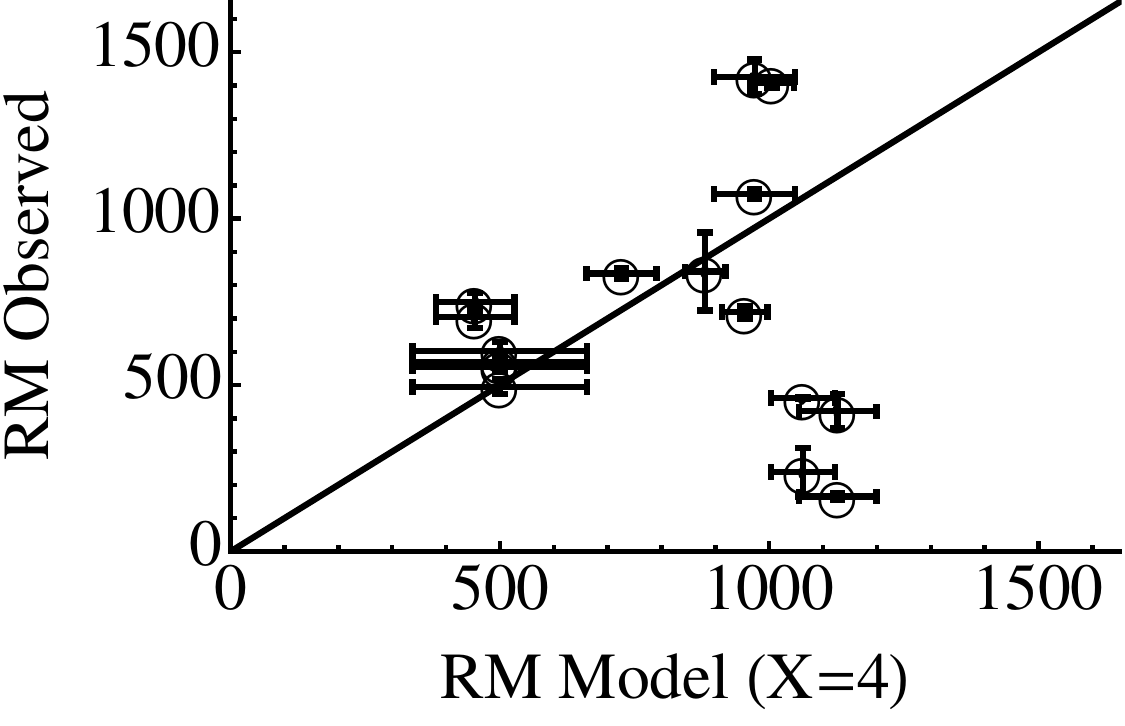}}\quad
\subfloat[\label{fig:HSRM}]{
\includegraphics[width=0.4\textwidth]{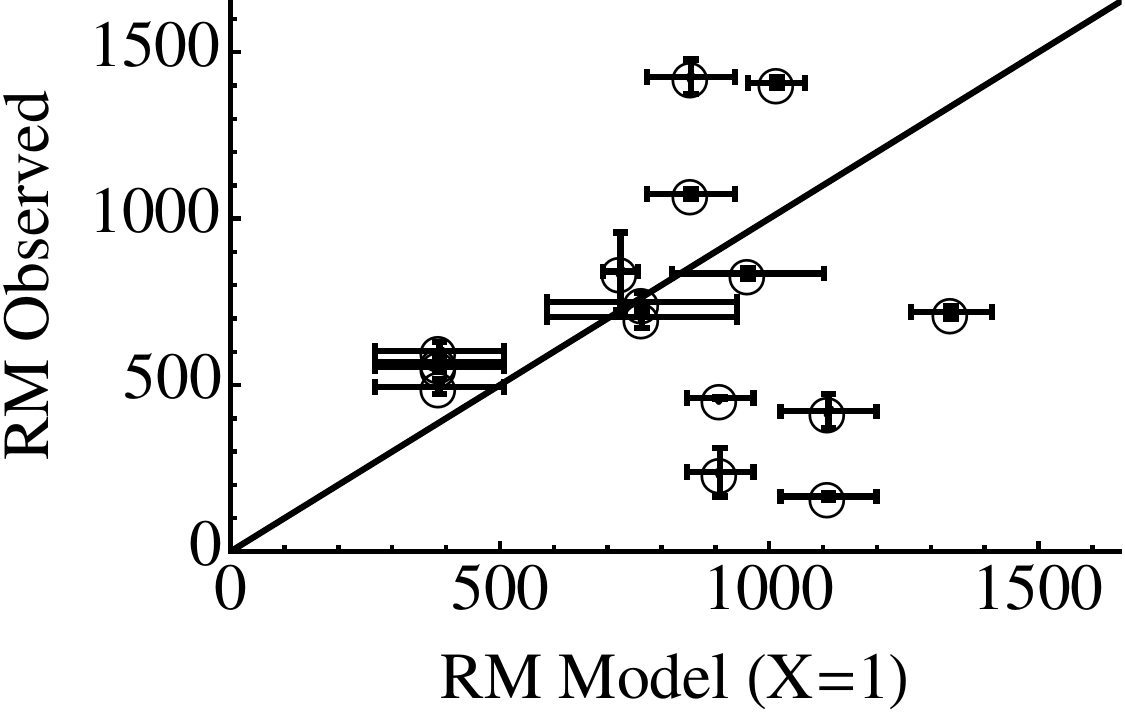}}
%
\caption[Comparison of Calculated RM and Observed RM]{Plots comparing the observed RM values (\radm) with calculated RM values using the local values of the shell radii and electron density for (a) Equation (\ref{RM3}) with $\theta$ = 71\ddeg and (b) Equation (\ref{RM2}) with $\theta$ = 33\ddeg. These are the peak values of $\theta$ from the Bayesian analysis (Section \ref{sec:bayjoe}). The  line shows the case of perfect agreement between the observed RM and the model RM values. Only the sources with $\xi$ $\leq$ R$_0$ are plotted (Table \ref{tab:localpars}).}
\label{fig:rmvrm}
\end{figure}

Figures \ref{fig:SRM} and \ref{fig:HSRM} plot the observed RM vs the modeled RM values calculated with Equations (\ref{RM3}) and (\ref{RM2}), respectively, for sources with $\xi$ $\leq$ R$_0$. The line in these figures represents the case of perfect agreement between the observed RM and calculated RM values. The errors for the modeled RM values are propagated from the fits of T$_B$, R$_0$, and R$_1$. For both cases, Figure \ref{fig:SRM} and Figure \ref{fig:HSRM}, there is rough agreement between observations and the model when the model predicts intermediate RM, i.e., +400 $\leq$ RM $\leq$ +1000 \radm. When the models predict high RM, however, the results are mixed. For some lines of sight, the observed RM are in reasonable agreement with the model values. However, in other cases, the observed RM is considerably lower.

It is unclear physically why this is the case. It is possible that some of these lines of sight pass through regions in which \textbf{B} is strongly modified from the simplified forms assumed in our analysis. In such cases, reversal of the shell magnetic field for some portion of the line of sight could lead to a drop in the total RM. Alternatively, it is possible that lines of sight characterized by low observed RM relative to model predictions are probing plasma cavities within the shell. This possibility is rendered less likely by the fact that the results of Figure \ref{fig:rmvrm} have taken into account the emission measure along a slice at the same position angle $\psi$ as the source. It has been suggested to us (D. Schnitzeler, private communication) that these low RMs for some lines of sight could indicate a low value of the filling factor $f$. This is more-or-less equivalent to the above suggestion that the anomalously low RMs are due to cavities in the nebula.

It is possible to have very high values of the RM that could potentially wrap within the bandwidth, which may cause depolarization. However, we do not believe this to be the case for the RM values we report. There are two reasons as to why we do not expect this: 1) the reduced $\chi^2$ analysis showed no deviations of $\chi$ from $\lambda^2$, and 2) in the RM Synthesis analysis, we are sensitive to up to $|\phi_{\textrm{{\scriptsize max}}}|\sim$ 3.0$\times$10$^{5}$ \radm, so we would be able to detect large RM values. In general, we do not believe that the observed RMs are the source of disagreement with the calculated values of the RM.

\subsubsection{Effect of Possible Filling Factor on Inferred Shell Parameters\label{sec:filling}} 

The analysis above interprets our observations in terms of highly simplified shell models. In all the models, the density is assumed uniform along the line of sight, and in the homogeneous models of Section 5.1, the density is constant at all points within the shell (illustrated in Figure 6 of \citealt{Whiting:2009}). If X = 1, the magnetic field is also uniform along the line of sight.  In the alternative case we have considered, X = 4, the field is constant with one value along half the line of sight (one half of the chord through the nebula), and constant with another value on the other half. This homogeneity has been assumed in recognition of the patent oversimplification of these models in describing the true \HII region associated with the Rosette Nebula.

Another parameter which frequently appears in the literature in discussions of \HII regions is the filling factor $f$, conventionally defined as the fraction of the \HII region volume that is occupied by ionized gas, as opposed to a vacuum which allegedly occupies the remainder of the volume.  \citet{Harvey:2011} adopt a value of $f = 0.1$ on the basis of numbers reported in the literature, and this value is incorporated in their estimates of $B_{\parallel}$. \citet{Purcell:2015} adopted a value of  $f = 0.3$ for the Gum Nebula, but noted that it was poorly constrained, and could be significantly larger.  Planck Collaboration (2015), in their study of the Rosette Nebula, assumed a value of  $f = 1$, apparently in view of the absence of data indicating a smaller number.

There are two ways of estimating the filling factor for an \HII region.
\begin{enumerate}
\item In the first investigation of this matter, \citet{Osterbrock:1959} measured the density in the central part of the Orion Nebula from the density-sensitive line ratio O{\sc II} 3729/3726.  They showed that the inferred density, if taken as uniform through the nebula, would result in an EM much larger than that obtained by measurement of the radio thermal continuum emission.  The resolution, illustrated in Figure 7 of \citet{Osterbrock:1959}, was to posit that the ionized gas was contained in clouds occupying a small fraction of the nebular volume.  This same approach, using density-sensitive infrared lines of S{\sc III} was used by \citet{Herter:1982} to reach similar conclusions about the \HII regions G75.84+0.4 and W3 IRS 1.  We were unable to find suitable measurements of the O{\sc II} 3729/3726 doublet in the Rosette Nebula, so an estimate of $f$ in this way could not be done.
\item A second method for determining the filling factor was presented in \citet{Kassim:1989}. It relies solely on radio measurements and could be termed a radiometric method of determining $f$.  The estimate of $f$ from \citet{Kassim:1989} results from requiring consistency between the optically thin radio flux of a nebula and estimates of EM obtained from the low frequency radio spectrum, at and below frequencies at which the nebula becomes optically thick (see Equations (6) and (7) of \citealt{Kassim:1989}).  The filling factor is then defined as
\begin{equation}
f \equiv \frac{\Omega}{\Omega_{\textrm{beam}}}
\end{equation}
where $\Omega$ is the inferred solid angle subtended by regions of emission, and $\Omega_{\textrm{beam}}$ is the beam of the radio telescope, i.e. the solid angle over which all emission is averaged.
\end{enumerate}
An apparently equivalent (i.e. using the same physical concepts) formulation of the \citet{Kassim:1989} method compares the measured brightness temperature $T_B$ of the \HII region in the optically thick part of the spectrum (more precisely, the antenna temperature $T_A$ attributable to the {\em thermal} emission of the nebula) to the independently-determined electron temperature $T_e$,
\begin{equation}
f =  \frac{\Omega}{\Omega_{beam}} = \frac{T_B}{T_e}.
\label{eq:filling}
\end{equation}
Equation (\ref{eq:filling}), applied to the data of \citet{Kassim:1989} on \HII regions associated with the S53 complex, reproduces the filling factors in Table 5 of \citet{Kassim:1989}.  It should be noted that the filling factor calculated in the method of \citet{Kassim:1989} will not usually equal that defined by \citet{Osterbrock:1959}, but should indicate if the latter is substantially less than unity, and may be used as a rough estimate in the absence of other information.

An estimate of $f$ using the method of \citet{Kassim:1989} is possible for the Rosette Nebula.  \citet{Graham:1982} assembled radio continuum measurements of the Rosette Nebula over a wide range of frequencies, and showed that the nebula becomes optically thick at a frequency of about 400 MHz.  From a fit to the lower frequency part of the spectrum, they obtain an optically-thick brightness (antenna) temperature of $T_B = 4100^{+700}_{-500}$ K.  \citet{Celnik:1986} reports an independent measurement of the electron temperature $T_e$, based on radio recombination line observations.  Celnik measures $T_e = 4700 \pm 600$ K if LTE level populations are assumed, and $T_e = 5800 \pm 700$ K on the basis of a non-LTE calculation.  Adopting the latter value for  $T_e$, a value of $f = 0.71 \pm 0.13$ results.  Given the uncertainty in this whole procedure, this value should probably not be considered strongly inconsistent with $f = 1$.

We now briefly discuss the consequences of a filling factor $f \sim 0.7$ on the analysis presented in this paper.  If $f$ (now taken to be the fraction of line of sight occupied by clumps containing plasma) is less than unity, the true density in those clumps will be larger by a factor of $1/\sqrt f$ than deduced for a uniform shell. Likewise, the value of $B_{0z}$ in the clumps will be larger by the same factor.  Since our modeling of RM observations constrains the quantity $B_{0z} = B_0 \cos \theta$ (Equation (\ref{eq:B})), the consequences of $f < 1$ are given in the following expression,
\begin{equation}
\frac{B_0^1 \cos \theta^1}{\sqrt f} = B_0^f \cos \theta^f 
\label{eq:bf}
\end{equation}
with the variables defined as follows.  The quantities $B_0^1$ and $ \theta^1$ are the magnetic field strength and inclination angle in the case $f=1$ (homogeneous shell).  These are the quantities presented in the previous sections.  The variables $B_0^f$ and $\theta^f$ represent the corresponding quantities when $f \leq 1$.  If we assume that $B_0^1 = B_0^f = B_0$ is known, Equation (\ref{eq:bf}) becomes an identity relating $\theta^1$ and $\theta^f$.

Table \ref{tab:filling} reports our results concerning the effects of a value of $f$ less than unity.  This table presents values of $\theta^f$ corresponding to different values of $f$.  Column 1 gives the value of the compression factor $X$, since this determines $\theta$.  Columns 2, 3, and 4 give the corresponding values of $\theta^f$ for $f$ = 1, 0.71,  and 0.50.  The value $\theta^{f = 1}$ gives the reference case for a nominally uniform shell.  The case  $\theta^{f = 0.71}$ corresponds to the estimate of $f$ based on a comparison between the brightness temperature and electron temperature for the Rosette.  Finally,  $\theta^{f=0.50}$ corresponds to a more strongly clumped case, intermediate between more extreme estimates of $f \leq 0.1$ obtained by \citet{Osterbrock:1959}, \citet{Herter:1982}, and \citet{Kassim:1989}, and the unclumped limit of $f = 1$.
\begin{table}
\centering
\caption{Results of Filling Factor\label{tab:filling}}
\begin{tabular}{cccc} \hline
  Model & $\theta^{f=1}$ & $\theta^{f = 0.71}$ & $\theta^{f=0.5}$ \\ \hline
  $X=4$ & $73^{\circ}$ & $70^{\circ}$ & $66^{\circ}$ \\
  $X=1$ & $47^{\circ}$ & $36^{\circ}$ & $15^{\circ}$ \\ \hline
\end{tabular}
\end{table}

The obvious effect of $f < 1$ is to require that the interstellar magnetic field $\vec{B}_0$ be closer to the line of sight.  This tendency is clearest for the $X=1, f = 0.50$ case.  To the extent that a larger value of $\theta$ indicates a more realistic model, given that the Rosette Nebula is roughly in the anticenter direction (\citealt{Savage:2013}, Section 4.1), smaller values of $f$ are less preferred.

This thinking can be pursued to obtain a highly model-dependent constraint on $f$. Obviously $\cos \theta^f \leq 1$, which is satisfied for the $X=1$ case for $f \geq 0.46$.  A smaller value of $f$ can be accommodated within the $X=4$ model.  However, in both cases the assumption of a filling factor significantly less than unity implies that the ISM magnetic field  $\vec{B}_0$ at the location of the Rosette Nebula would be significantly rotated from its mean value for an azimuthal magnetic field. A final remark is that less highly inclined fields (larger $\theta$) can be accommodated if the true value of $B_0$, the magnitude of the interstellar magnetic field at the location of the Rosette Nebula, is larger than our assumed value of 4 $\mu$G. 
\subsubsection{Bayesian Analysis for Inhomogeneous Shell Models \label{sec:bayjoe}}
In Section \ref{sec:joe}, we marginally relaxed the homogeneous shell model to account for inhomogeneity in the shell of the Rosette Nebula. We employ the Bayesian analysis to determine which shell model better describes the observed RM values and whether the interstellar magnetic field is alternatively strongly modified or unmodified in the shell. We perform the Bayesian analysis in the same manner described in Section \ref{sec:bay}. Rows three and four of Table \ref{tab:alpha} lists the results of the first iteration of the Bayesian analysis with $\alpha$ = 0.2 and the second iteration where $\alpha$ = $\sigma_{\textrm{SD}}$ = 0.43. In Figures \ref{fig:local1} and \ref{fig:local2}, the integrand of Equation (\ref{eq:pD}),  p($\theta$$\mid$M,I)p(D$\mid$M,$\theta$,I), is plotted as a function of $\theta$ for the model with amplification of the magnetic field and that without, respectively. The calculated values of the RM for the inhomogeneous models are listed in columns 7 and 8 in Table \ref{tab:localpars} using the values of \thetaP~ from the second iteration of the Bayesian analysis.

Using the localized parameters for the shell and electron density, $\alpha$ = 0.43, and N = 16, p(D$\mid$M$_1$,I) = 9.2$\times$10$^{-46}$ with $\theta_{\textrm{{\scriptsize peak}}}$ = 71\ddeg, and p(D$\mid$M$_2$,I) = 2.6$\times$10$^{-45}$ with $\theta_{\textrm{{\scriptsize peak}}}$ = 33\ddeg. The Bayes factor is B$_{12}$ = 0.37 (B$_{12}^{-1}$ = 2.7) and \(|\ln{B}| = 1.0\), which is ``weak at best'' \citep{Gordon:2007} evidence on the Jeffreys' scale in favor of the model of \citet{Harvey:2011} model. The conclusions of the Bayesian analysis for the inhomogeneous case verges on neither model being favored. While incorporating inhomogeneity by determining the electron density and inner and outer radii for each line of sight yields  better and more accurate models, this procedure does not eliminate large residuals between the models and the measurements.  These large residuals apparently prevent the Bayesian analysis from convincingly selecting one of the models as the preferred one.

\subsection{Summary of Bayesian Test of Models}
The Bayesian analysis was undertaken to determine which of the simplified models discussed in Section \ref{sec:emmodels} is a better representation of the Rosette Nebula Faraday rotation data. The results of the Bayesian analysis are listed in Table \ref{tab:alpha}. If a global shell model (i.e., the same values of R$_0$, R$_1$, and n$_e$ for all lines of sight) is applied to the data, the Bayes factor is B$_{12}$ = 0.19 (B$_{12}^{-1}$ = 5.4) and \(|\ln{B}| = 1.7\), indicating that the picture of \citet{Harvey:2011}, i.e., a model without modification of \textbf{B}$_0$, is weakly favored in its ability to reproduce the observations (Section \ref{sec:bay}). 

If an attempt is made to accommodate the inhomogeneity of the nebula by having the effective values of R$_0$, R$_1$, and n$_e$ change for different slices through the nebula, the Bayes factor is B$_{12}$ = 0.37 (B$_{12}^{-1}$ = 2.7) and \(|\ln{B}| = 1.0\), which is ``weak at best'' \citep{Gordon:2007} in favor of the \citet{Harvey:2011} model. 

The results of the Bayesian analysis depend on the choice of  $\alpha$.  This probably is a consequence of the fact that as  $\alpha$ increases, the role of outliers in the likelihood function decreases.  In any case, as the value of  $\alpha$ is increased from 0.20 to 0.43, the preferred inhomogeneous model changes from weakly in favor of the X=4 model to weakly in favor of the X=1 (unmodified) model.

Our analysis is also sensitive to our choice of prior, i.e., allowing $\theta$ to range on a surface of a hemisphere. This choice of prior can bias the results at higher values of $\theta$. We initially performed the Bayesian analysis with a uniform prior such that \(p(\theta|M,I)=\frac{1}{\Delta\theta}, \) where $\Delta\theta$=$\frac{\pi}{2}$. However, the choice of prior does not change which model is favored in the Bayesian analysis.

The conclusions of the Bayesian analysis are not strong, and the inability of this analysis to identify a clear preferred model is apparently due to the oversimplified nature of all models considered.  Inhomogeneities or other structures in the real nebula partially mask the average properties of the nebula.  However, a positive point is that either model can reproduce the magnitude and extent of enhanced RMs over the right angular extent on the sky, for plausible values of the magnitude and orientation of the interstellar magnetic field. For the global models, which predict that the RM should depend only on the impact parameter $\xi$, the plot to consult is Figure \ref{fig:2model}. Both models account for the abrupt onset of the RM at the edge of the nebula, and both reproduce, with plausible values of the independent parameter $\theta$, the values of RM actually observed. However, an equally striking aspect of Figure \ref{fig:2model} is the large dispersion of measured RMs about the model curves. Apparently, variations in the plasma structure of the stellar bubble, describable by variations in n$_e$, $\mid$B$\mid$, and $\theta$ are as important as the description of the mean shell characteristics, and in fact obscure the signature of this mean structure.

For the case of the ``inhomogeneous models'', the relevant plot is Figure \ref{fig:rmvrm}. In the inhomogeneous case, the predicted RM does not depend only on $\xi$, and we make a comparison between the model and observable values for all lines of sight. Here, again, the models meet with some success in that they can reproduce the magnitude of the RMs observed, and are also able to \textit{partially} account for the variation of RM from one line  of sight to another. That is, the models seem to successfully predict the upper envelope of the observed RMs. However, the models do not account for the fact that some lines of sight, expected to have large RMs, have intermediate to low values. Again we must appeal to spatial variations within the nebula, potentially describable as large-scale turbulence, that cause variations in the values of RM.

As a final comment, it is worthwhile to know that plausible variations in one parameter, $\theta$, allow two models that significantly differ in a physical sense to reasonably account for the data. We suspect that this statement would extend to the model presented by \citet{Planck:2015}. Their physical description of the nebular shell is very different and probably more realistic than those considered in this paper. Their result for the prediction of RM($\xi$), presented in Figure 6 of their paper, is similar in form to that shown in Figure \ref{fig:2model} of our paper, with different but plausible values of $\mid$B$_0$$\mid$ and $\theta$. However, it is not capable of accounting for substantial dispersion in RM for lines of sight with the same value of $\xi$.

\section{Summary and Conclusions\label{sec:con}}
\begin{enumerate}
\item We performed polarimetric observations using the VLA of 11 new radio sources observed through the shell of the Rosette Nebula. This significantly increases the number of lines of sight probed relative to that of \citet{Savage:2013}.

\item We obtained RM measurements for 15 lines of sight (including secondary components), using two techniques: a least-squares fit to \chilam~and RM Synthesis. The measured RM values are in excellent agreement between the two techniques. We measure an excess RM of +40 to +1200 \radm~due to the shell of the Rosette Nebula and confirm the background RM of $+$147 \radm~from \citet{Savage:2013}. The large range in values of RM through the shell is an indicator of inhomogeneity in the shell, which is beyond the scope of the present paper.
\item We employ a Bayesian analysis to determine which of two models better represents the observed dependence of RM with distance from the center of the nebula, a model with a shock-enhanced magnetic field in the shell or one without. We first treat the Rosette Nebula as a spherically-symmetric shell having global parameters for the inner and outer radii of the shell and the electron density. With this calculation, the model without modification of the interstellar field is favored. The  Bayes factor is 5.4 (B$_{12}$$^{-1}$) and on the Jeffreys' scale, the analysis of \citet{Harvey:2011} is weakly favored.
\item We perform a second Bayesian analysis using localized parameters determined from radio continuum data to account for the inhomogeneity of the shell. In this analysis, we model the inner and outer shell radii and calculate the electron density to each line of sight. 
We then employ these localized parameters in the Bayesian analysis and find the model without modification of the magnetic field in the shell is weakly favored. The Bayes factor is 2.7, which on the Jeffreys' scale ($| \ln{\textrm{B}} |$ = 1.0) is considered ``weak at best'' \citep{Gordon:2007}.
\item The results of the Bayesian analysis are dependent on the value of $\alpha$, which specifies the expected departure of the measurements from the model.  For a value of $\alpha \simeq 0.4$, neither model is significantly superior to the other in a Bayesian sense.  We attribute this to the fact that neither model can account for large departures of the observed RMs from the model predictions.  This situation is unlikely to change, even for analytic and numerical models that include more sophisticated physics than the empirical expressions we have employed. 
\item Regardless of the specific choice of model, the magnitude and spatial extent of the rotation measure ``anomaly'' is completely consistent with independent information on the plasma density in the nebula, and the interstellar magnetic field.
\item To further investigate  the role of magnetic fields in \HII regions, one would want more lines of sight through the shell of the Rosette \HII region and the stellar bubble to obtain a large sample of RM values. In many cases though, such a sample is limited by the number of strong, linearly polarized, extragalactic sources behind the nebula. Simulations of \HII regions and stellar bubbles with magnetic fields, such as those by \citet{Stil:2009}, \citet{vanMarle:2015}, \citet{Ferriere:1991}, and \citet{Walch:2015}, are vital to understanding how the magnetic field and the density contribute to the large RM values within the shell of the nebula. It would be of considerable interest to compare calculated RMs on lines of sight through these simulations with the measured  values.  Such comparison would indicate if those simulations are capable of reproducing the average features of Faraday rotation through these ``Faraday Rotation Anomalies'' \citep{Whiting:2009}, as well as the large fluctuations in the observed RMs with respect to the expected values.   Finally, similar analyses with more \HII regions and stellar bubbles would further illuminate the role of the magnetic field and would allow one to consider differences in stellar composition, wind luminosities, and ages of the nebulae. Presently, we are undertaking such an analysis as presented in this paper for W4/IC 1805 and IC 1396. 
\end{enumerate}
\acknowledgments{This research was supported at the University of Iowa by grants AST09-07911 and ATM09-56901 from the National Science Foundation. We sincerely thank Dr. Wolfgang Reich of the Max Planck Institut f\"ur Radioastronomie for sending us the archival 4.95 GHz image of the Rosette Nebula made with the Effelsberg radio telescope. These data permitted the analysis of Section \ref{sec:joe}. We would also like to acknowledge T. Robishaw (Dominion Radio Astrophysical Observatory (DRAO), Canada) for use of his IDL code used in our RM Synthesis analysis, and Dominic Schnitzeler of the Max Planck Institue f\"ur Radioastronomie for reviewing this paper and providing us with excellent comments. Finally, we thank the referee of this paper for a helpful and collegial review.}
\bibliography{FRbib,LBVbib,compbib,FollowupBib}
\bibliographystyle{apj}

\end{document}

%% file: defs.tex
\newcommand{\thetaP}{$\theta_{\textrm{{\scriptsize peak}}}$}

\newcommand{\chilam}{$\chi(\lambda^{2})$}

\newcommand{\HII}{H\,{\sc ii} } 
\newcommand{\ddeg}{\ensuremath{^{\circ}}} 
\newcommand{\Msun}{\ensuremath{\textrm{M}_{\odot} }} 

\newcommand{\kms}{km s$^{-1}$}
\newcommand{\infinity}{\ensuremath{\infty}}
\newcommand{\radm}{rad m$^{-2}$}